\documentclass[12pt,preprint,xdvi]{aastex}
\usepackage{graphicx}
\usepackage{psfig}

\newcommand{\be}{\begin{equation}}
\newcommand{\ee}{\end{equation}}
\newcommand\eq{eq.}
\newcommand\eqs{eqs.}
\newcommand\fig{Figure}
\newcommand\figs{Figures}

\def\bvec{{\bf B}}

\def\vvec{{\bf v}}

\def\xvec{{\bf x}}

\def\xhat{{\bf \hat{x}}}
\def\yhat{{\bf \hat{y}}}
\def\zhat{{\bf \hat{z}}}

\def\half{\hbox{${1\over2}$}}
\def\third{\hbox{${1\over3}$}}

\def\avg#1{{\bigl\langle#1\bigr\rangle}}
\def\bpprime{B'_{\perp}}
\def\tlife{\tau_{\rm plug}}
\def\etal{{\em et al.}}
\def\ie{{\em i.e.}}
\def\lgl{LGL09}
\def\arcsec{$^{\prime\prime}$}
\def\la{\mathrel{\hbox{\rlap{\hbox{\lower4pt\hbox{$\sim$}}}\hbox{$<$}}}}
\def\ga{\mathrel{\hbox{\rlap{\hbox{\lower4pt\hbox{$\sim$}}}\hbox{$>$}}}}
\def\natexlab#1{{#1}}

\begin{document}

\title{A Quantitative Model of Energy Release and Heating by Time-dependent, 
Localized Reconnection in a Flare with a Thermal Loop-top X-ray Source}

\author{D.W. Longcope,$^{1,\dagger}$ 
A.C. Des {J}ardins,$^1$
T. Carranza-Fulmer,$^{2,3}$ J. Qiu,$^1$}
\affil{$^1$ Department of Physics, Montana State University,
  Bozeman, Montana 59717\\
$^2$ Montana State University, Solar Physics REU program\\
$^3$ City University of New York, New York, NY\\
$^{\dagger}$ e-mail: dana@solar.physics.montana.edu}



\begin{abstract}
We present a quantitative model of the magnetic energy stored and then released through magnetic reconnection for a flare on 26 Feb 2004.  This flare, well observed by RHESSI and TRACE, shows evidence of non-thermal electrons only for a brief, early phase.  Throughout the main period of energy release there is a super-hot ($T\ga30$ MK) plasma emitting thermal bremsstrahlung atop the flare loops.  Our model describes the heating and compression of such a source by localized, transient magnetic reconnection.   It is a three-dimensional generalization of the Petschek model whereby Alfv\'en-speed retraction following reconnection drives supersonic inflows parallel to the field lines, which form shocks heating, compressing, and confining a loop-top plasma plug.  The confining inflows provide longer life than a freely-expanding or conductively-cooling plasma of similar size and temperature.  Superposition of successive transient episodes of localized reconnection across a current sheet produces an apparently persistent, localized source of high-temperature emission.  The temperature of the source decreases smoothly on a time scale consistent with observations, far longer than the cooling time of a single plug.  Built from  a disordered collection of small plugs, the source need not have the coherent jet-like structure predicted by steady-state reconnection models.  This new model predicts temperatures and emission  measure consistent with the observations of 26 Feb 2004.  Furthermore, the total energy released by the flare is found to be roughly consistent with that predicted by the model.  Only a small fraction of the energy released appears in the super-hot source at any one time, but roughly a quarter of the flare energy is thermalized by the reconnection shocks over the course of the flare.  All energy is presumed to ultimately appear in the lower-temperature $T\la20$ MK, post-flare loops.  The number, size and early appearance of these loops in TRACE's 171 \AA\ band is consistent with the type of transient reconnection assumed in the model.
\end{abstract}


\section{Introduction}

Since the introduction of magnetic reconnection
\cite{Giovanelli1947,Sweet1958,Parker1957,Petschek1964}, 
many models have proposed it as an 
element of flare-related phenomena, 
such as chromospheric ribbons 
\cite{Carmichael1964,Sturrock1968,Hirayama1974,Kopp1976}, associated coronal mass ejections \cite{Moore1980,Forbes1995}, above-the-loop X-ray sources 
\cite{Masuda1994,Tsuneta1997} and supra-arcade downflows \cite{McKenzie1999,McKenzie2000,Sheeley2004}.

The term ``magnetic reconnection'' is commonly used to refer to either or both of two different effects: topological changes to magnetic field lines and conversion of magnetic energy into other forms.  Only the first effect, topological change, requires a parallel electric field and thereby qualifies as ``magnetic reconnection'' in the strictest sense.  It has been successfully quantified by several kinds of observation.  The amount of flux being topologically changed can be computed by tracking the ribbons across photospheric field in a flare \cite{Forbes1984,Poletto1986,Qiu2002}, computing the flux ejected by CMEs 
\cite{Qiu2007,Longcope2007}, or measuring the fluxes of retracting downflows 
\cite{McKenzie2009}. Energy conversion, on the other hand, can occur together with the topological change, as in the diffusion-dominated Sweet-Parker models 
\cite{Parker1957,Sweet1958} or as a subsequent response to it, as in the Petschek model. The latter case would be more properly termed ``post-reconnection energy conversion'',
but Petschek's model is customarily referred to as one of reconnection.

Recent theoretical progress has revealed that in order to be genuinely fast, the reconnection electric field must be localized to a small portion of a current sheet 
\cite{Birn2001,Biskamp2001,Kulsrud2001}.  Since such a small region will contain negligible magnetic energy, energy conversion must occur {\em away} from the region of flux transfer, as a post-reconnection {\em response} to small-scale topological change.  This requirement of physical separation between the non-ideal electric field and the energy conversion offers the hope that energy conversion can be modeled without direct knowledge of, or appeal to, the specific physics dictating flux transfer.

In Petschek's prototypical model of fast magnetic reconnection, energy conversion occurs at slow magnetosonic shocks (SMSs) originating in the small flux-transfer region 
\cite{Petschek1964,Sonnerup1970,Vasyliunas1975,Soward1982}.  While these models invoked a resistive electric field to transfer the flux, it has since been established that any localized process will produce the same external response provided the plasma may be treated by fluid equations (and thereby support collisional shocks) on its largest scales 
\cite{Erkaev2000}.  The two-dimensional, steady-state model has also been generalized to transient reconnection \cite{Biernat1987,Nitta2001}, reconnection between skewed magnetic fields \cite[\ie\ with a magnetic field component, or ``guide field'', in the ignorable direction, sometimes called ``two-and-a-half'' 
dimensions] {Petschek1967,Soward1982b,Skender2003} and finally to transient 
reconnection in three dimensions \cite{LGL09}.  We find that all such models predict very similar energetics since they all model external fluid responses to an assumed flux transfer of negligible dimension.

Tsuneta (\citeyear{Tsuneta1996}) presented a thorough application of Petschek's reconnection model to observations of a solar flare.  He demonstrated that the temperature structure observed above the long-duration event on 21 Feb 1992 was consistent with the SMSs of two-dimensional Petschek reconnection.  He showed further that local energy fluxes were consistent with the model, but could not, for a number of reasons, turn this into a global accounting of net energy released.  First of all, the flare took place on the eastern solar limb making a reliable model of the coronal magnetic field difficult.  Indeed, invoking Petschek's original two-dimensional model tacitly neglects any possible magnetic-field component along the line of sight.  Secondly, the reconnecting current sheet was probably a dynamical after-effect of a CME whose energetic contribution could overwhelm that of the flare itself.  Finally, the impulsive phases of most large flares appear to include non-thermal particles, which fluid-based reconnection models cannot easily incorporate, but which account for a significant energy \cite{Lin1971,Strong1984,Emslie2004} .

Except for the adverse geometry, the difficulties faced by Tsuneta (\citeyear{Tsuneta1996}) are likely to be common to most large flares.  Non-thermal particles pose an especially difficult challenge for global energetics.  While much progress has been made understanding the mechanism for the their acceleration \cite[for reviews]{Miller1997,Aschwanden2002b}, most models pre-suppose a non-ideal electric field, fluid turbulence, or plasma waves.  Large-scale fluid models of reconnection are consistent, to some extent, with each of these, but do not include coupling to non-thermal particles.  Since these particles ultimately account for a significant part of the released energy their effect is almost certain to be non-negligible. 

A small subset of large flares show little evidence for non-thermal particles, but instead exhibit hard X-ray (HXR) thermal bremsstrahlung from a super-hot plasma ($T\ga30$ MK).  Rare as they are, these flares offer a unique opportunity to comprehensively quantify the energy release in a flare using a fluid reconnection model.  Furthermore, these flares are often of the so-called {\em compact} variety and thus lack CMEs which could otherwise muddy the energy estimate.  Lacking an eruption, the pre-flare magnetic field can be assumed to be near equilibrium and its energy more easily estimated.  Such an ideal flare occurred on 26 Feb 2004 and was well observed by both RHESSI and TRACE at high cadence with little interruption.  This flare has been found to be one of the rare (one in ten) 
X-class flares unassociated with any CME \cite{Wang2007}.
We present here a model of how magnetic reconnection produced the phenomena observed in this flare, including estimates of the entire energy release process.
  
Spectrally resolved HXR images show a compact source probably located atop the flaring loops.  This source is assigned an electron density which is high but typical of flaring plasma, ($n_{\rm e}\ga10^{11}\,{\rm cm}^{-3}$).  The high density is required by both the large observed X-ray flux and in order that electrons of such high energies to be collisionaly thermalized.   High-temperature loop-top sources such as this have been observed before \cite{Acton1992,Petrosian2002,Jiang2006}, and have proven particularly challenging to explain.  We believe that shocks from fast reconnection offer a promising avenue for modeling the source, since they would both compress and heat the plasma. 

The source's persistence time vastly exceeds the time that it would take for the localized structure to conductively cool or to freely expand.  Some models have addressed the former difficulty by proposing that a turbulent reduction of the electron thermal conductivity permits the high-temperature plasma to cool more slowly \cite{Jiang2006}.  This does not explain, however, how its high pressures are confined to the top of the loop.  A second approach is
to assume the observed, persistent structure is actually a
super-position of sequential transient events; this approach is common in models of 
flaring loops \cite{Hori1997,Reeves2002,Warren2002,Warren2006}.  The
temperature of these features does, however, change smoothly and
gradually, in a manner resembling a slow cooling.  Proposing that it
consists of multiple isolated elements then requires an explanation
for their smooth collective behavior.  This is presumably a signature
of the evolving energy conversion process --- possibly magnetic reconnection.

In this work we show that magnetic reconnection occurring sporadically in small patches across a pre-flare current sheet will produce a super-position of high-density, 
high-temperature sources consistent with the observations on 26 Feb 2004.  Moreover, the values of the temperature and emission measure, as well as their gradual temporal evolution, is consistent with the model.  The specific physics responsible for transferring flux within the reconnection patches does not enter into these predictions and cannot therefore be constrained by the observations.  The mean rate of transfer, which is to say the number of new patches per unit time, does enter, but this is a global quantity that can be measured \cite{Forbes1984,Poletto1986,Qiu2002}.  While this kind of disordered process is less obviously related to Petschek's model than a long, steady outflow jet, we show that it is consistent with the observed size and temperature of the loop-top source.

This combination of observations and model are presented as follows.
The next section describes the observational data of the flare.  Section 3 then presents a model of the pre-flare magnetic field, including the geometry of the three-dimensional current sheet on which magnetic reconnection subsequently occurs.  We then present, in section 4, a general model of transient, localized magnetic reconnection within a current sheet.  The Alfv\'en speed and angular discontinuity characterizing the pre-reconnection current sheet directly predict the temperature of the reconnection outflow.  We show that this prediction is common across all models of  fast reconnection: steady state or transient, two-and-a-half or three-dimensional.  In section 5 we apply this model to the observations of 26 Feb 2004.   We maintain that evidence in this particular flare favors a transient reconnection model.  We predict an emission measure that can be directly compared to observation.  We use different means to measure the mean flux-transfer rate and find that all agree and predict flaring emission measure comparable to the observation.  Section 6 uses the reconnection to model the full flare, including an account of its energetics.  Finally, section 7 discusses the possible applicability of the new model to a broader class of flares.

\section{The Flare: 26 Feb 2004}

\subsection{Active region 10564 -- emergence}

\begin{figure}[htp]
\psfig{file=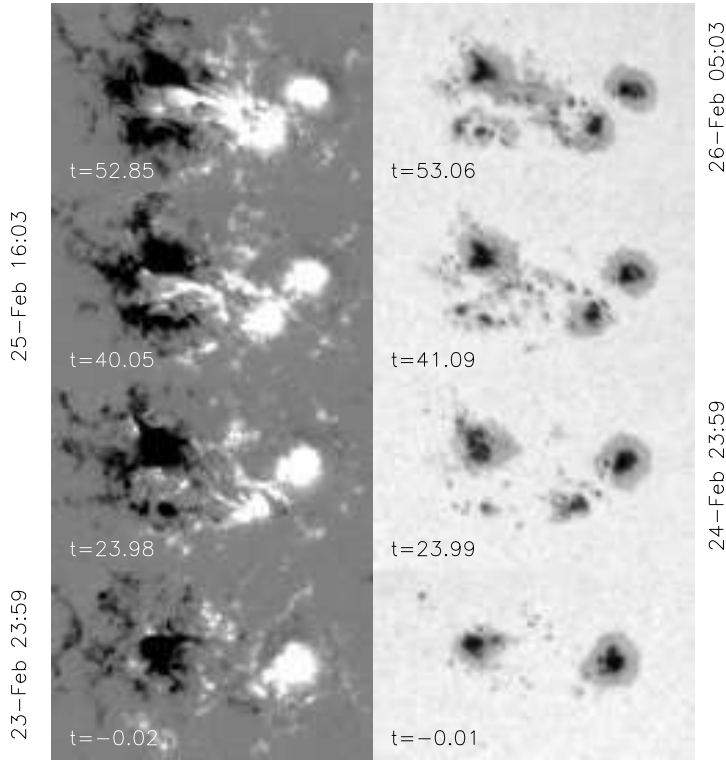,width=4.8in}
\caption{MDI data of new flux emergence in AR 10564.  Rows are data 
from four successive times, progressing upward, spanning 54 hours.  Printed 
on each panel is the time since 00:00UT on 26 Feb 2004.  The left
column is the line-of-sight magnetogram scaled between $\pm1000$ G; 
the right is the continuum image.}
	\label{fig:mdi_sum}
\end{figure}

The 26 Feb 2004 flare (SOL2004-02-26T02:03:00L161C076)
occurred within the complicated, but basically
bipolar, active region 10564.  The simple bipolar active region had one 
umbra of each polarity until, on 24 Feb, new
flux began to emerge into it.  This new flux eventually
developed its own umbrae and the new positive and old negative
penumbrae merged (see \fig\ \ref{fig:mdi_sum}).  After roughly 50
hours of steady emergence an X-class flare occurred within the
region.

\begin{figure}[htp]
\psfig{file=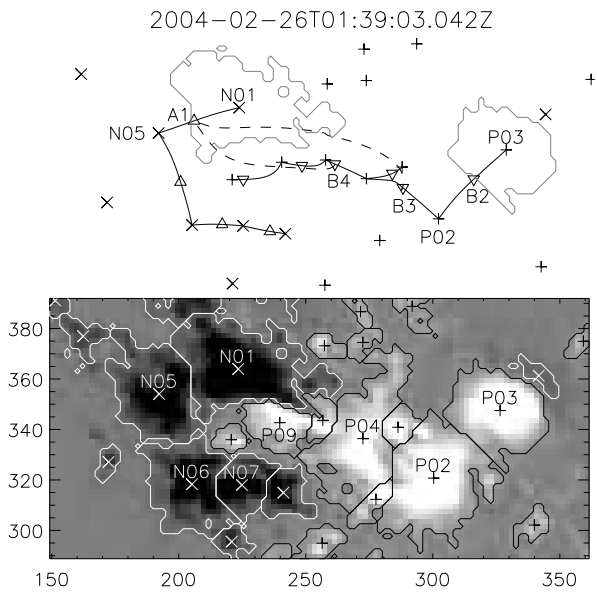,width=4.8in}
\caption{Partition of the line-of-sight magnetogram from just before the
flare.  The lower panel shows the magnetogram, scaled to $\pm1000$
G.  Lines denote the boundaries of a partition, and $+$ and $\times$
are the centroids of each region.  The upper panel shows the same
centroids along with null points (triangles) and the principal
spines (solid) connecting them.  The dashed curves are separators
connecting null point $A1$ to $B3$ and $B4$.}
	\label{fig:mdi_fp}
\end{figure}

To distinguish between new and old flux we partition the line of sight
MDI magnetogram as shown in \fig\ \ref{fig:mdi_fp}.  The radial
component, $B_r$, is derived from the line-of-sight measurement
assuming the field to be perfectly radial.\footnote{To assess the accuracy of the radial-field assumption \fig\ \ref{fig:mdi_sum} has a longitude axis and marks showing the time the old polarities cross central meridian.  Any systematic error would lead to an artificial center-to-limb variation centered at the $\times$.}  Those
pixels exceeding $|B_r|\ge B_{\rm thr}=125$ G are grouped into unipolar
regions according to the horizontal gradient in $B_r$
\cite{Barnes2005,Longcope2009d}.  The regions labeled $N01$ and $P03$
constitute the old bipole and all other partitions are from emergence
since the beginning of 24 Feb.

\begin{figure}[htp]
\psfig{file=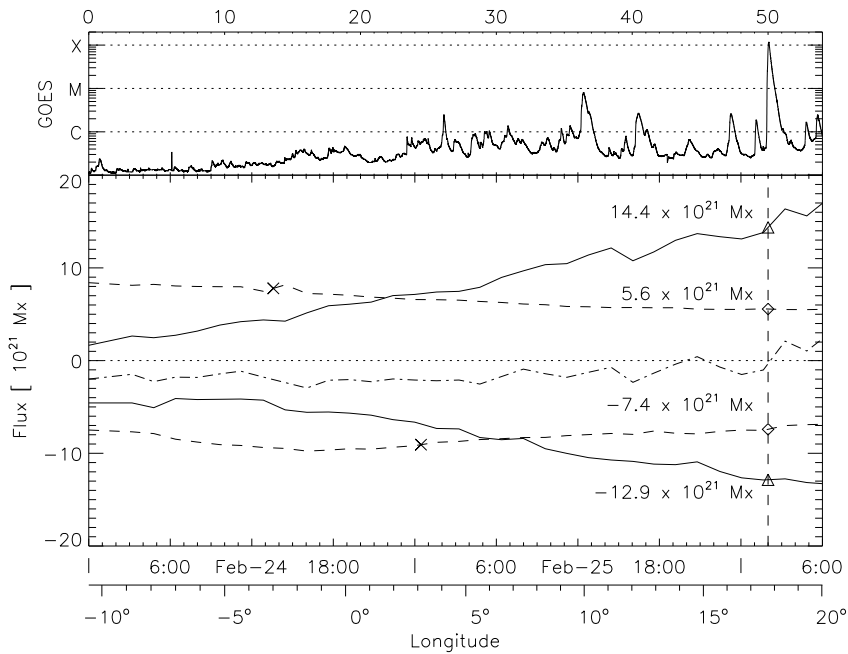,width=4.8in}
\caption{The fluxes in the new (solid) and old (dashed) polarities 
of AR 10564; the broken line is the total signed flux of all regions.
The value at the time of the flare, 26 Feb 2:03UT
(vertical dashed line), is 
given and indicated with a symbol.  The signed sum of all fluxes, shown as
a broken line, is close to zero.  
The times at which each of the old polarities crosses
central meridian is shown with an $\times$, and the longitude of their
mid-point is shown by the bottom axis.  
The top panel shows the GOES 1\AA\ -- 8\AA\ flux.  The top axis gives
hours past 24 Feb 00:00.}
	\label{fig:em_flux}
\end{figure}

The net flux and centroid of each
region  ($\Phi_i$ and $\bar{\xvec}_i$) is found by integrating its surface
field.  The time evolutions of old flux, $\Phi_{P03}$ (positive) and
$\Phi_{N01}$ (negative), are shown as dashed curves in \fig\
\ref{fig:em_flux}.  At the beginning of the data each has approximately
$8\times10^{21}$ Mx of flux.  Over the course of our data set the
fluxes in each polarity gradually decreases until they are
$\Phi_{P03}=5.6\times10^{21}$ Mx and 
$\Phi_{N01}=7.4\times10^{21}$ Mx at the time of the flare.
The new flux, comprising the remainder of the
regions in the partition, increases much more rapidly over this period
reaching twice that of the old flux ($|\Phi|\simeq13\times10^{21}$ Mx)
by the time of the flare.  The emergence is steady and there is no
clear feature in the evolution at the time of the flare (vertical
dashed line).  There is a small  excess of negative flux (broken line) due to the 
omission of positive flux either outside the field of view or below the threshold strength, 
$B_{\rm thr}=125$ G.

We interpret flare observations in terms of a magnetic model built using these photospheric observations combined with assumptions about flux emergence.  The partitioning of the photospheric field described above makes no assumption about the state of the coronal field, but does provide a framework for quantifying its connectivity \cite{Longcope2002b}.
Regardless of the state of the coronal field, every coronal field line can be assigned to a domain based on the photospheric regions of its footpoints: its connectivity.  Different coronal fields have different connectivity and the level of difference between two possible fields is quantified by the differences in domain fluxes \cite{Longcope2009d}.   We use the potential field extrapolated from the pre-flare magnetogram of 1:39 (\fig\ \ref{fig:mdi_fp}) as a reference for constructing such differences.   The potential field has the minimum possible magnetic energy so the degree by which a given field differs from it relates to the amount for free energy in the field \cite{Longcope2001b,Longcope2004}.

Provided the post-flare coronal field has lower magnetic energy than the pre-flare field, we expect its connectivity to be closer to that of the potential field.  The potential field connectivity is characterized in terms of the magnetic skeleton 
\cite{Priest1997,Longcope2002b,Beveridge2005}, illustrated by the top panel of 
\fig\ \ref{fig:mdi_fp}.  This stylized figure shows sources ($+$s and $\times$s) and several of the magnetic null points (triangles) of the potential field.  Solid curves show the spine lines from each null point, and dashed lines are two separators which extend into the corona from null points $A1$, $B3$ and $B4$.  The spine lines trace out the edges of separatrix surfaces which divide the coronal magnetic field into different domains.  Most significant for the present case, they divide the pre-existing coronal field, anchored to $P03$ and $N01$, from newly emerging field.

The pre-flare field will differ significantly from the potential field since it must contain significant currents.  Rather than try to measure these currents we infer them from the connectivity after assuming that no significant reconnection occurred prior to the flare.  This means that in the pre-flare field the old photospheric regions, $P03$ and $N01$, remain unconnected to the others, such as $P09$ or $N07$, which recently emerged.  The potential field, and therefore the post-flare field, contains significant connections between new and old regions regions $P09$ and $N01$, due to their proximity.  By our flux-emergence assumption, however, this connection contains zero flux in the pre-flare field.  It is differences such as this that lead to large free energies in the pre-flare field.  The forging of new connections through reconnection will reduce this difference thereby releasing stored magnetic energy and presumably powering the flare.

\subsection{The flare loops}

\begin{figure}[htp]
\psfig{file=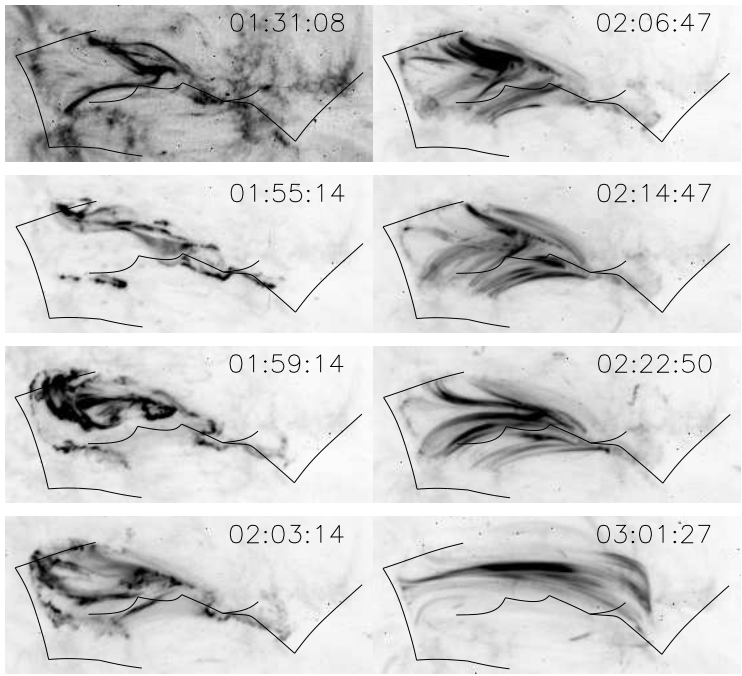,width=4.8in}
\caption{TRACE 171\AA\ images from times before (1:31) and during the
flare, shown in inverse greyscale.  The grey scale is enhanced by
a factor of eight in the first image and by $4/3$ in the final three
(2:14 -- 3:01).  Dark lines in each frame show the spines from the
magnetic skeleton from \fig\ \ref{fig:mdi_fp}.}
	\label{fig:tr_sum}
\end{figure}

EUV images provide evidence supporting our assumption that flare-related reconnection creates field lines with new connectivity.
TRACE obtained 171\AA\ images at a 30-second cadence throughout the flare.
Figure \ref{fig:tr_sum} shows an image from before the flare (1:31) and seven
subsequent images during the flare.  Post-flare loops 
occupy the region between by the spine curves, corresponding 
to a coronal volume beneath the associated separatrices.
In images from early in the flare the 171\AA\ emission is
primarily chromospheric, including moss, footpoints and flare ribbons
(e.g.\ 1:55:14).  Following a common pattern 
\cite{Gorbachev1988,Longcope2007}, the flare 
ribbons follow the topological spines. 
Coronal loops are visible as early as 1:57 and become the primary
features by 2:04:14.

\begin{figure}[htp]
\psfig{file=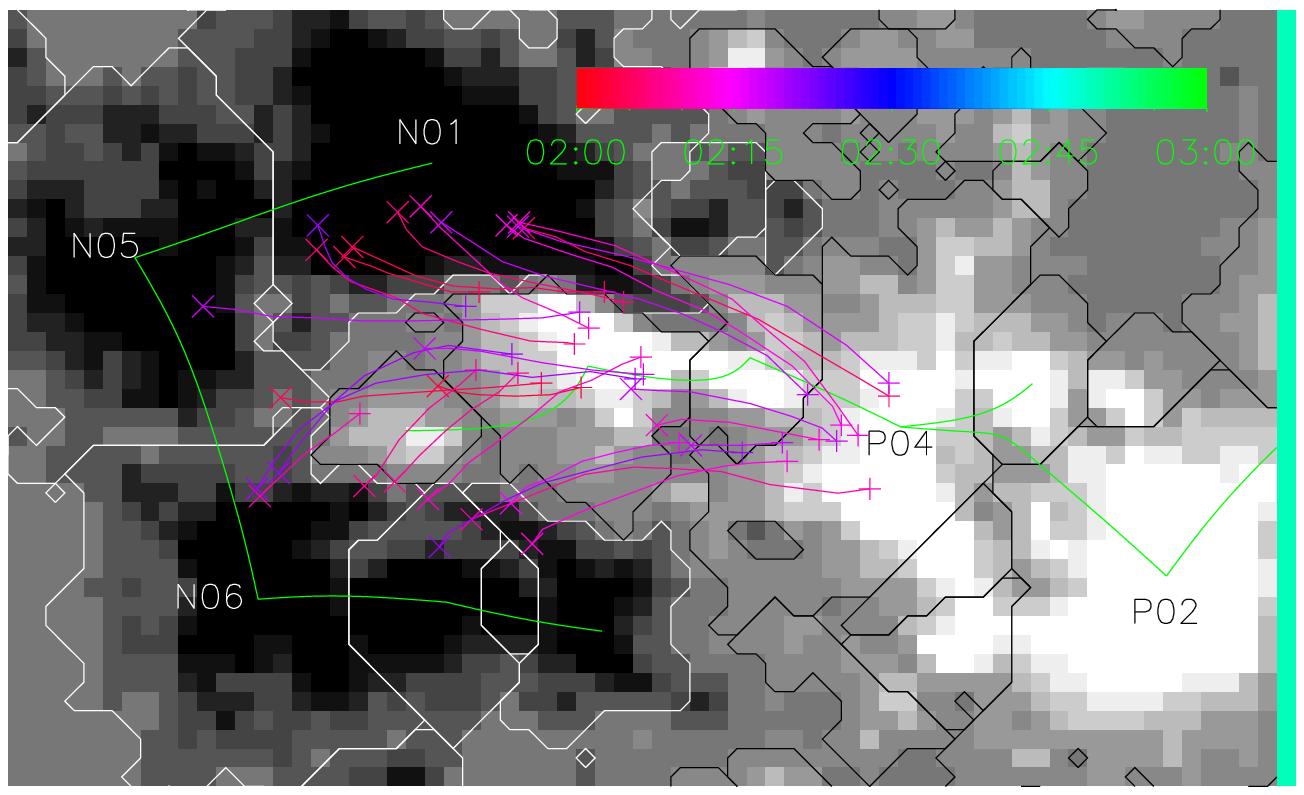,width=4.0in}
\psfig{file=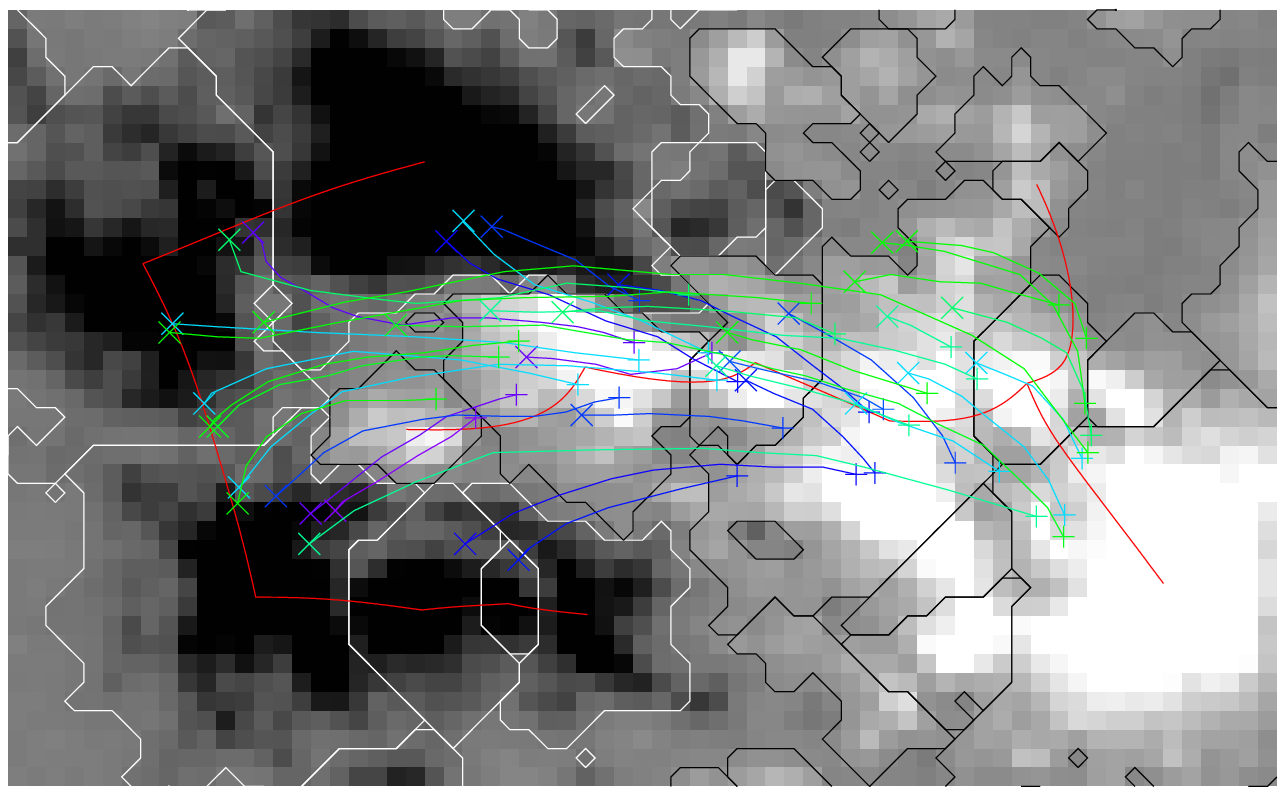,width=4.0in}
\psfig{file=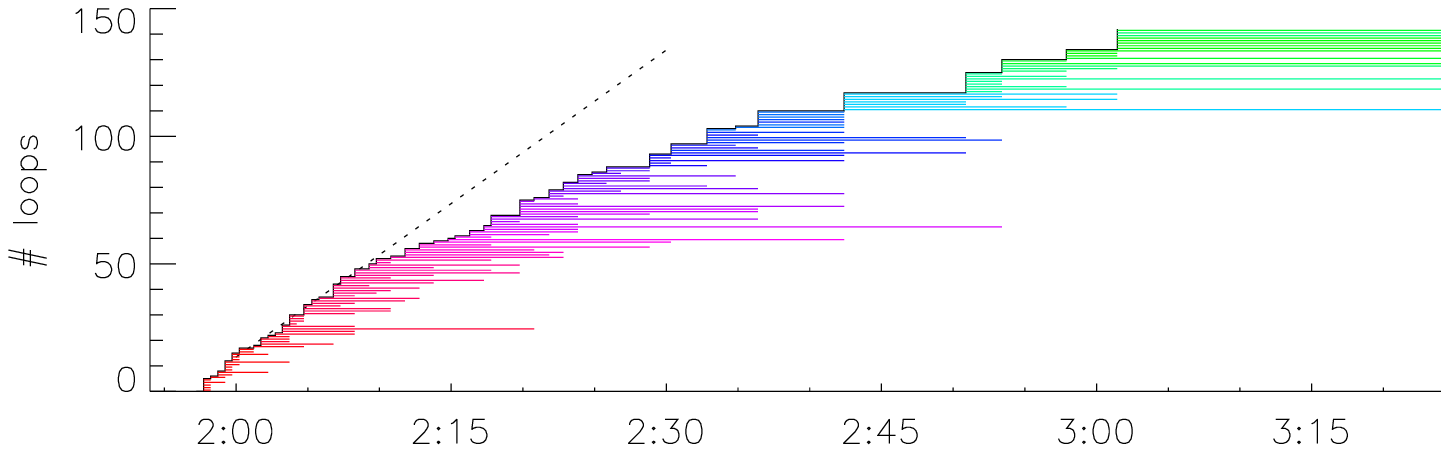,width=4.0in}
\caption{Coronal loops visible in TRACE 171\AA\ images are shown as
color curves atop the partitioned MDI magnetogram.  Colors 
from red to green represent times form 2:00 to 3:00UT, as indicated in
the top color bar.  The upper panel shows the loops which appear
before 2:18, and the middle panel shows later loops.  For clarity,
only those loops
visible for longer than the median lifetime are shown.
The bottom panel show the accumulated number of loops visible before
any time.  Horizontal bars show the lifetime of each loop, colored
with the same code as the loops themselves.}
	\label{fig:loops}
\end{figure}

The visible coronal loops are summarized in \fig\
\ref{fig:loops}.  143 loops were visually identified and manually
traced in the collected 30-second cadence 171\AA\ TRACE images.    Those
loops persisting longer than the median lifetime, 6.0 minutes, are
plotted over the 
partitioned magnetogram with color indicating the time of first
appearance.  Early, short loops connect new positive regions $P09$ and $P04$
to the old nearby region $N01$.  Later in the flare, longer connections are
made between $N01$ and the more distant region $P02$.  This is still
newer than region $P03$ which was part of the original bipole.  The apparent 
lengths of these loops increase with time approximately linearly according to the 
empirical relation
\be
  L_{\rm EUV}(t) ~=~\dot{L}(t-t_0) + L_0~~,
  	\label{eq:Lloop}
\ee
where $\dot{L}=4.7$ Mm hour$^{-1}$ and $L_0=6.3$ Mm using 
$t_0=$2:00:00.  At the
onset of the flare new loops were appearing at a rate of four per
minute, as indicated by the dotted line in the bottom panel.

\begin{figure}[htp]
\psfig{file=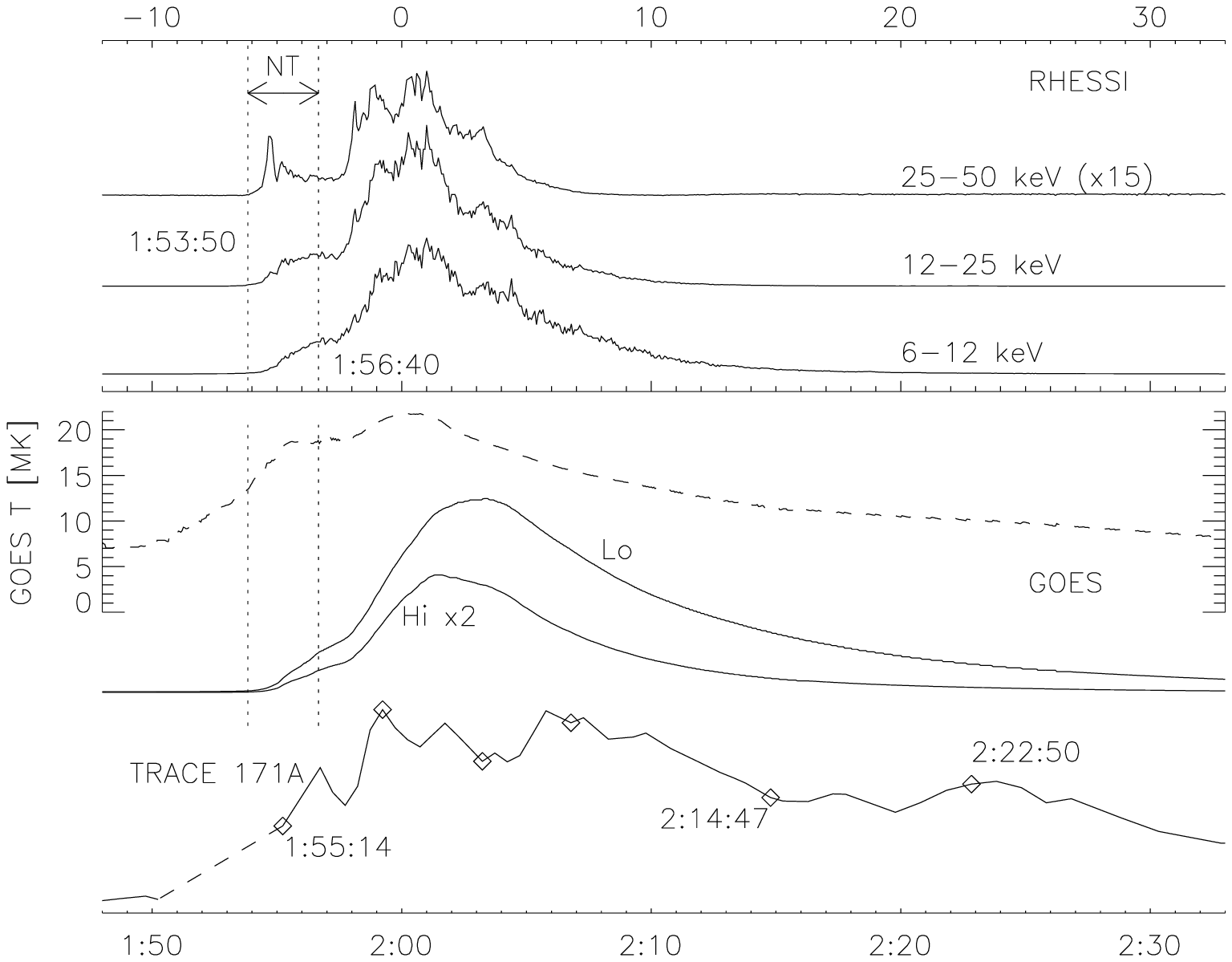,width=4.8in}
\caption{A composite light curve of the flare on 26 Feb 2004.  
All curves show intensity on a linear scale.
The bottom curve is the integrated
intensity of TRACE 171\AA\ from the region shown in \fig\
\ref{fig:tr_sum}.  The zero level is suppressed, but the
level prior to 1:50 shows the pre-flare background.
The dashed section is a 5-minute data gap and
diamonds show the times at which the central six panels of \fig\
\ref{fig:tr_sum}.  Above
this are the curves from the low energy (Lo, 1 -- 8\AA) and high
energy (Hi, 0.5 -- 4\AA) channels of GOES; the latter has been
multiplied by two to facilitate comparison.  The dashed curve above
them is the plasma temperature derived from the ratio of GOES energy
channels \protect\cite{Garcia1994}.  Top curves are integrated intensity
from RHESSI channels, from bottom to top, 6 -- 12, 12 -- 25 and 
12 -- 25 keV.  Vertical dotted lines show the beginning
and end of the non-thermal phase of the flare.}
	\label{fig:ltcrv}
\end{figure}

The overall time history of the flare is summarized in the composite
light curve of \fig\ \ref{fig:ltcrv}.  The first evidence of
the flare is an impulsive brightening, at 1:53:50, in high energy 
X-rays (RHESSI 25 -- 50 keV channel at top).
The lower energy X-ray channels begin rising at about 
this same time, as do both of
the GOES X-ray channels.  The first 171\AA\ image after a five-minute
gap (1:55:14) has total emission 
clearly elevated from the pre-flare level.  Three
minutes after the initial burst, at 1:56:40, all X-ray channels
rise more rapidly to a higher level.  It is during the times following 
this second rise that TRACE 171\AA\ images are dominated by loops.

\subsection{Super-hot thermal loop-top source.}

\begin{figure}[htp]
\centerline{\psfig{file=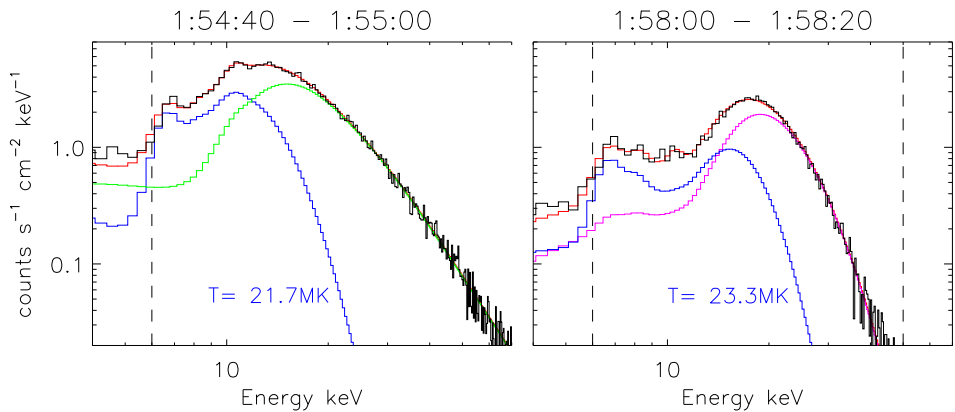,width=4.8in}}
\caption{RHESSI spectra from detector 3 integrated over 20 seconds plotted in black and multi-component fits.   Separate components of each fit are plotted in different colors, and their sum is the red curve.  The fit was performed over the region delineated by the vertical dashed lines.  (left) An early phase in the flare, centered at 1:54:50, fit by one thermal component ($T=22$ MK, blue) and  thick-target  emission from a non-thermal component with power-law spectral index $\delta=6.4$ (green).  (right) A later phase in the
flare centered at 1:58:10 fit by two thermal components:$T=46$ MK (magneta) and
$T=23$ MK (blue).}
	\label{fig:specs}
\end{figure}

Hard X-ray emission from the flare occurs in two phases with distinct
spectral forms typified by the examples shown 
in \fig\ \ref{fig:specs}.  Each spectrum is a 20-second
integration of RHESSI detector 3.
Plasma properties such as temperature and emission measure are found by fitting the spectra with the {\sf OSPEX} software package \cite{Schwartz1996}. In determining the best-fit model for each spectrum examined here, the following considerations were taken into account: spectral analysis from a single detector at a time, individual background subtraction for each energy range (6 -- 12, 12 -- 25, 25 -- 50, 50 -- 100 keV), albedo correction, the presence of instrumental lines at ~8.5 and ~10.5 keV (not shown in the figure), modifications necessary due to individual detector response differences and pulse pile-up and the attenuator change from A1 (thin only) to A3 (thick and thin) at $\approx$1:57 and back to A1 at $\approx$2:07.

For every 20 sec. time interval from 01:54 -- 02:10, we modeled spectra in two distinct ways.  First we fit a thermal bremsstrahlung plus a non-thermal thick target (power law).  Next we did a separate fit to two thermal bremsstrahlung components (a two-temperature fit). We found that the first phase of the flare (approximately 01:53:50 -- 01:56:40) is best characterized by the first fit, a thermal plus non-thermal spectral model; we refer to this phase of the flare as the non-thermal phase. The second phase (approximately 01:56:40 -- 02:10) is best fit by the two temperature model with no non-thermal contribution; we refer to this phase as the thermal phase. During the thermal phase, the higher energy end of the spectrum (T $\ga$ 12 keV) is consistently better fit by a super-hot thermal bremsstrahlung component than by a thick target power law component. In the 01:58:10 spectrum in \fig\ \ref{fig:specs}, the super hot component (magneta) has a temperature of 46 MK

\begin{figure}[htp]
\centerline{\psfig{file=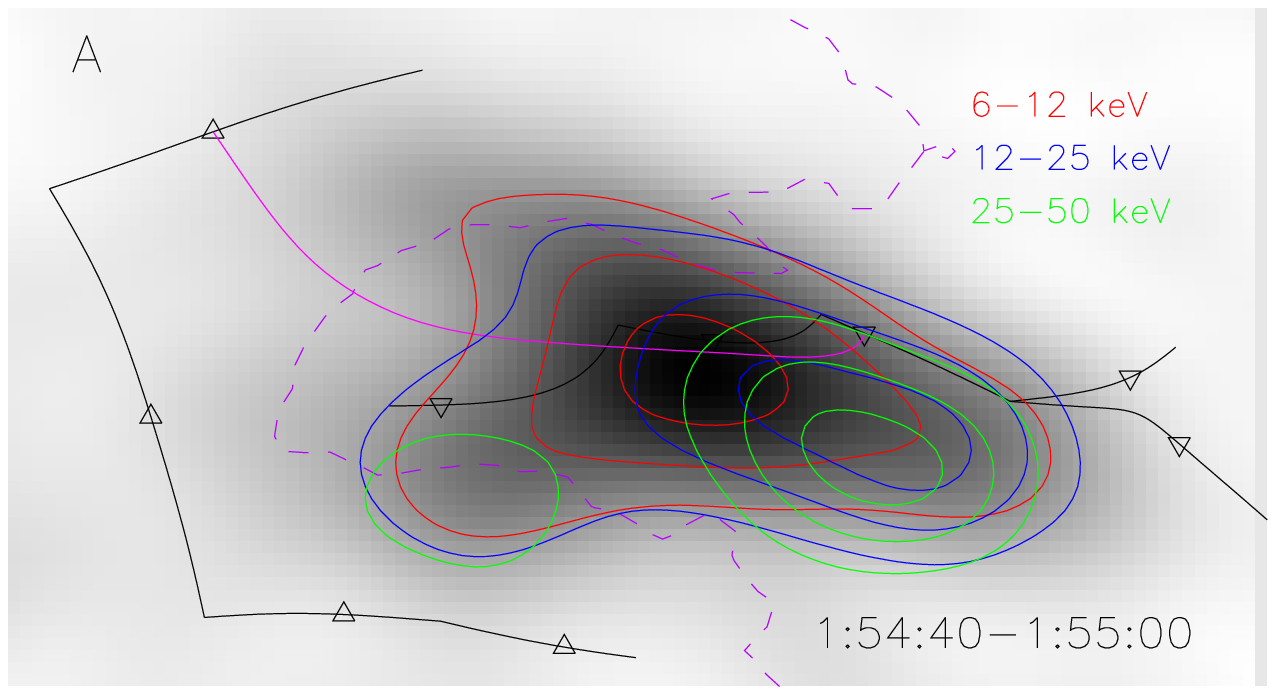,width=2.6in}\psfig{file=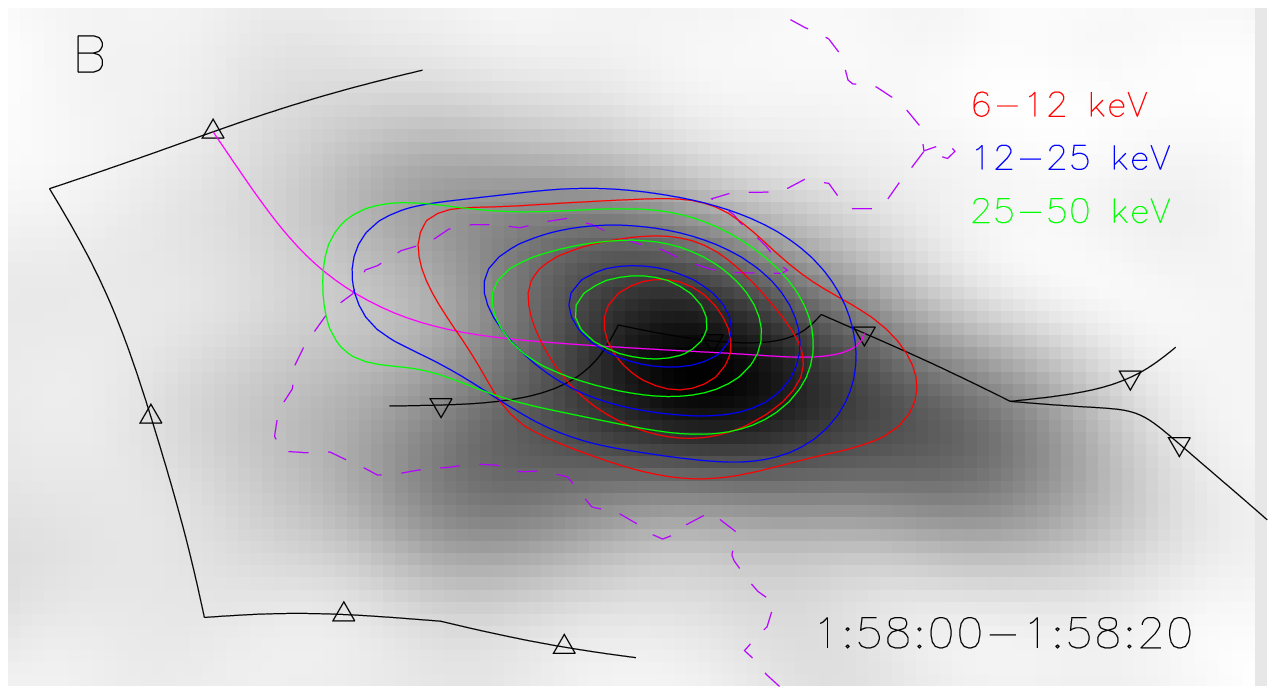,width=2.6in}}
\centerline{\psfig{file=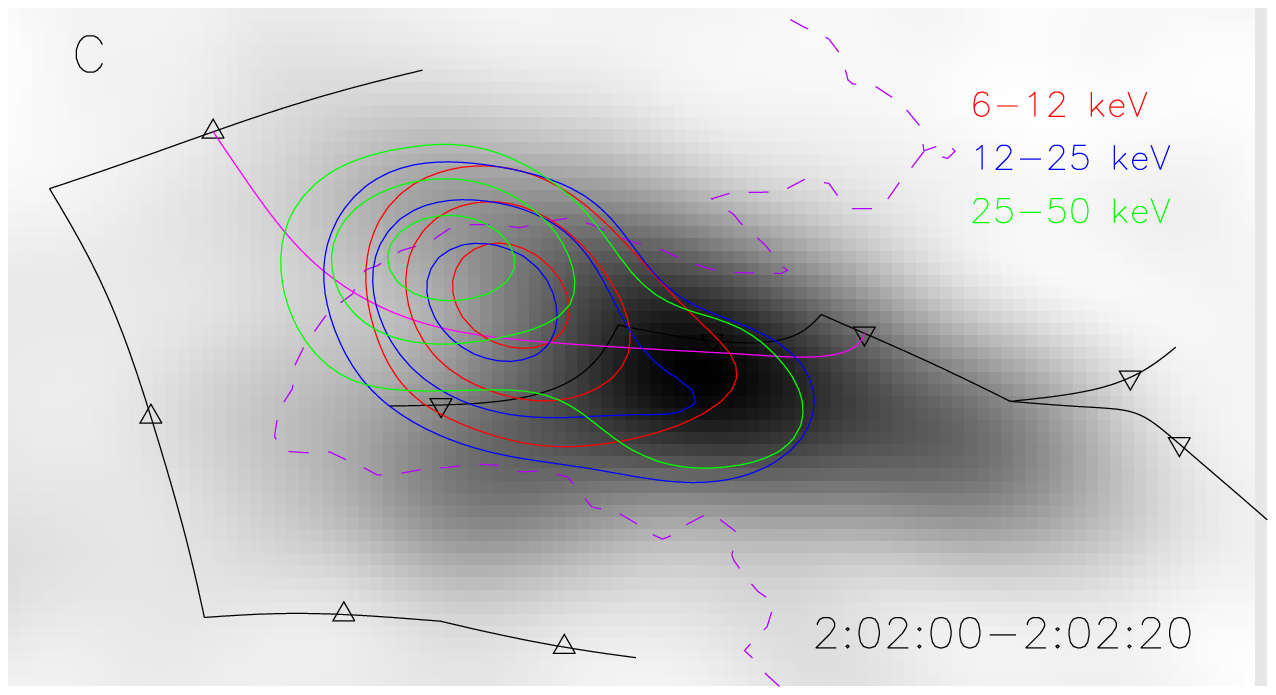,width=2.6in}\psfig{file=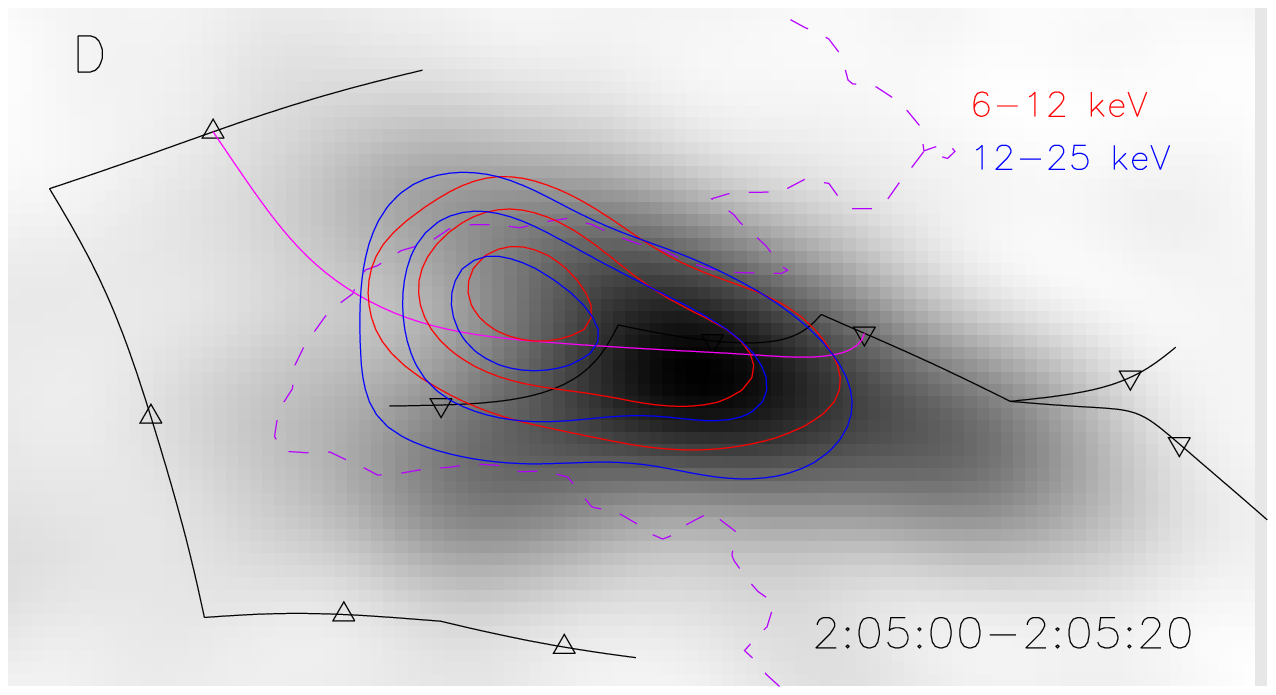,width=2.6in}}
\caption{RHESSI images formed using the CLEAN algorithm on 20-sec 
integrations of multiple detectors.  The grey scale 
in all images is from low-energy channel (6 -- 12 keV) integrated beginning at 1:54:00.
Contours are at levels of $50\%$, $70\%$ and $90\%$ of maximum for energy levels
6 -- 12 keV (red), 12 -- 25 keV (blue) and 25 -- 50 keV (green).  The panels show integrations
beginning at A: 1:54:40, B: 1:58:00, C: 2:02:00 and
D: 2:05:00.   Also shown are the 
spines (black), null points (triangles), one separator (magenta) and polarity inversion 
line (violet dashed).}
	\label{fig:rh_img}
\end{figure}

The  spatial morphology of X-ray emission also differs between the two phases, as illustrated by the RHESSI CLEAN images shown in \fig\ \ref{fig:rh_img}.  During the non-thermal phase (\fig\ \ref{fig:rh_img}a) the highest energy emission (25 -- 50 keV, green) originates in two distinct sources located on opposite sides of the PIL.  These are presumably from non-thermal particles striking the chromospheric footpoints of the reconnected field lines, perhaps linking $P09$ -- $N06$.  While the lower energy emission (6 -- 25 keV, red and blue) extends over the footpoints, it is more concentrated between the two sources.  This point is presumably closer to the apex of the loop, or the site of the reconnection.  It is notable that the peak of the 6 -- 12 keV emission (red) occurs near the separator field line (magenta).

During the thermal phase (\figs\ \ref{fig:rh_img}b -- d) the high and low energy emission 
originate from a single source near the separator field line.  It is elongated approximately parallel to the separator, with length $\approx16$ Mm and width $\approx9$ Mm.  Based on the magnetic geometry we believe this source is at the top of the post-flare loop system; it is a loop-top source viewed from above.  By 2:05 (\fig\ \ref{fig:rh_img}d) the count rate in the highest energy channel (25 -- 50 keV) had fallen too low to permit accurate imaging.  

\begin{figure}[htp]
\psfig{file=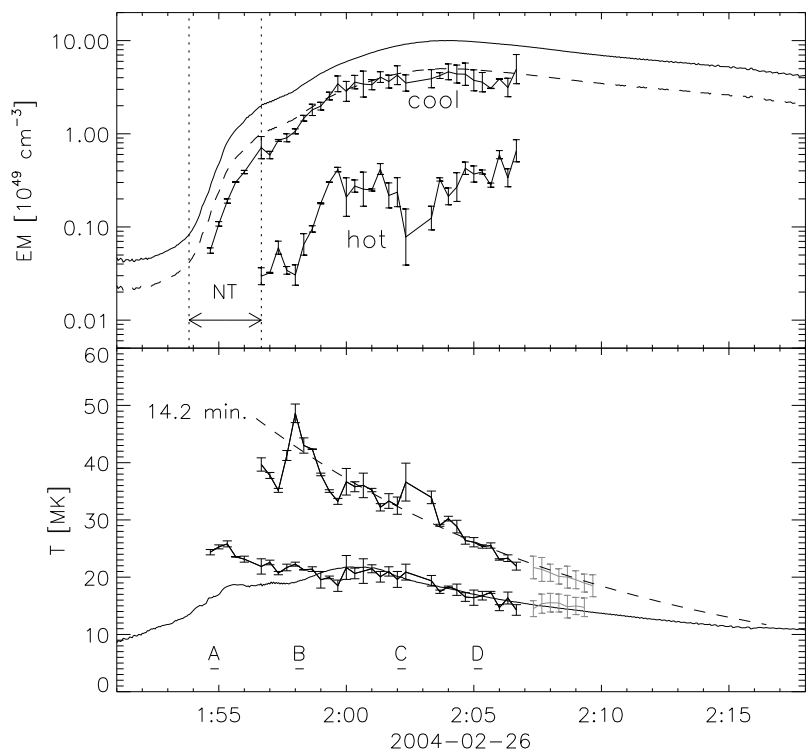,width=4.8in}
\caption{The temperatures and emission measures of the two thermal
components from detectors 3 and 4.  Curves show the simple average of
temperatures from the two detectors (bottom) and geometric means of emission
measures (top).  Error bars show the values from each detector.  The
smoother solid curves are the values derived from the two GOES
channels.  The dashed curve is the GOES-derived emission measure
divided by a factor or 2, after Hannah \etal\ 
(\protect\citeyear{Hannah2008}).  Lighter points show times after
2:07, where the two components were more difficult to discriminate.
The dashed temperature curve is an exponential fit to the super-hot
component; its decay time is $14.2$ minutes.  The imaging intervals, 
A -- D, from \fig\ \ref{fig:rh_img} are indicated along the bottom.}
	\label{fig:RHESSI_T}
\end{figure}

Spectra were formed from  both detectors 
3 and 4 over each 20-second interval after 1:54:40.  
All spectra resemble one of  the typical cases shown in \fig\ \ref{fig:specs}, but with different fitting parameters.  Time histories of the best-fit temperatures and emission measures of each
component are shown in \fig\ \ref{fig:RHESSI_T}.  
Spectra at times before 1:56:40 have non-thermal (NT) components and are best fit by a thick target power law and one thermal component whose temperature and emission measure are plotted alone.
All later spectra are fit by two temperatures.  By 2:07:00 the two temperatures are 
sufficiently close that it 
becomes difficult to discriminate between the components.  As a result, the fits from the different detectors give notably different values as indicated by larger error bars.  As early as 2:05:00,  the time of image \ref{fig:rh_img}d, the temperature of super-hot component had fallen sufficiently low (26 MK) that it does not contribute enough signal  to the 25 -- 50 keV 
channel to permit accurate imaging; for this reason there is no green contour in that image.

The super-hot component reaches as high as $T=48$ MK at the beginning.  It decreases
on an approximately exponential decay (dashed curve) with a time
constant of $14.2$ minutes, while the cool component decays even more slowly. It is notable that the cool component matches the
properties derived from the two GOES channels (after subtracting a background level determined from an interval prior to 1:00); the GOES-derived emission measure appears to be larger than the RHESSI value by roughly
a factor of 2 \cite{Hannah2008}.


\section{Model of the pre-flare magnetic field}

The picture which emerges from the foregoing data is of a sudden burst of
reconnection between the old bipole and the newly emerged flux.  In the
seminal model of such reconnection, proposed by Heyvaerts \etal\ 
(\citeyear{Heyvaerts1977}),
a current sheet forms between the two flux systems.  Reconnection
across this sheet releases the energy which powers the flare.  A
potential field above that configuration would contain a single coronal null
point from which four separatrices divide the coronal flux into four
domains: new flux, old flux and two mixed old/new domains.  
Remaining potential during emergence requires 
reconnection across the null point in order to convert new and old
flux into the mixed type field lines.  Without reconnection a current
sheet forms at the erstwhile null point.

The Heyvaerts \etal\ (\citeyear{Heyvaerts1977}) 
model is two-dimensional and the
emerging field is purely anti-parallel to the overlying field.
In the present case, however, the emerging flux is largely parallel to
the existing bipole $P03$ -- $N01$.  This leads to a modified 
three-dimensional model in which the potential field contains a separator
rather than a coronal null point.  A potential field extrapolated from
the magnetogram of \fig\ \ref{fig:mdi_fp} contains no coronal nulls
but many separators; two separators are shown in the upper panel.
Each separator is a magnetic field line ($\bvec\ne0$) lying at the
intersection of two separatrices.  Together these separatrices separate four different field line connectivities and the separator lies between all four.

\begin{figure}[htp]
\psfig{file=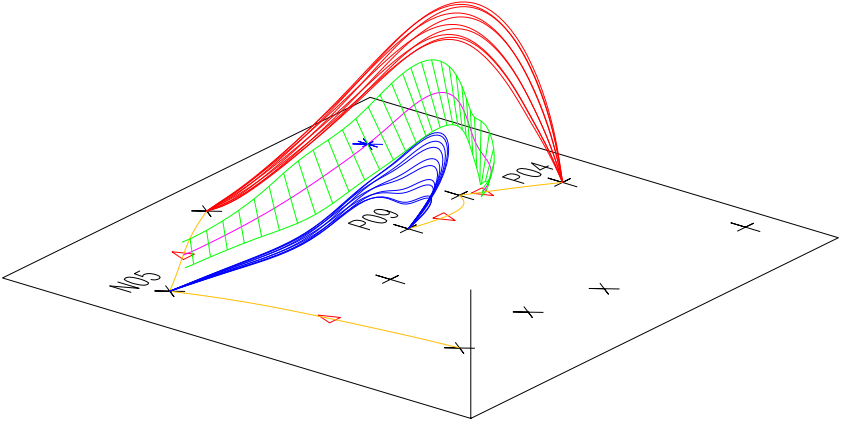,width=4.0in}
\caption{A perspective view of a current ribbon along the separator
field line connecting null points $B4$ and $A1$ (magenta).  The
current sheet is indicated by green ribs and the current flows
roughly parallel to the magenta separator.  Examples of post
reconnection field lines are indicated by red ($P04$ -- $N01$) and blue
($P09$ -- $N05$). A blue asterisk shows a point along the separator
located $z=7.9$ Mm above the photosphere.}
	\label{fig:cr}
\end{figure}

In analogy to the two-dimensional model of Heyvaerts \etal\ 
(\citeyear{Heyvaerts1977}),
emergence without reconnection leads to a current ribbon approximately
following the path of the potential-field separator.  Figure
\ref{fig:cr} shows the configuration of such a current ribbon along
the separator connecting photospheric null points $A1$ and $B4$.  There
is a component of magnetic field parallel to the current, often
called a {\em guide field}.  In spite of its three-dimensional
structure the equilibrium
structure of this current ribbon can be approximated using the 
two-dimensional model
of Green (\citeyear{Green1965}) and Syrovatskii 
(\citeyear{Syrovatskii1971}).

To analyze the current ribbon,
consider a plane perpendicular to the separator field line at one 
single point, such as the blue asterisk in \fig\ \ref{fig:cr}.  
Take this as the origin of the plane with local
coordinates $y$ and $z$.  In the neighborhood of the origin
the potential magnetic field within this plane is
\be
  \bvec^{\rm (v)}(y,z) ~=~ B_g\xhat ~-~ \bpprime( y\zhat + z\yhat )
  ~~,
	\label{eq:B_xpoint}
\ee
where $\bpprime$ is the local field gradient at the origin
\cite{Longcope1998,Longcope2004}.  At the coronal point indicated in
\fig\ \ref{fig:cr}, $B_g=331$ G and $\bpprime=34\,{\rm G\,Mm^{-1}}$.

The plane's coordinates have been
oriented so the field is radial along diagonals $45^{\circ}$ 
from the coordinate axes.  Unlike the two-dimensional case, these
diagonal lines cannot be associated with the field's separatrices,
since those are global features not derivable from local
properties of the field.  

\begin{figure}[htp]
\psfig{file=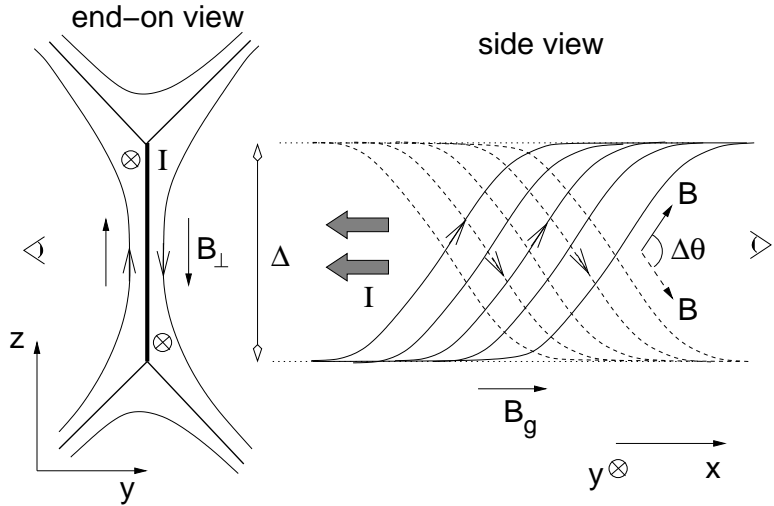,width=4.0in}
\caption{The geometry of a current sheet with a guide field
component.  The left panel shows the conventional ``end-on view'' 
of the sheet, along its current $-\xhat$.  Two Y-type neutral 
points, separated by $\Delta$ define the ends of the sheet.
Viewing this from the
left provides the less-conventional ``side view'', shown in the right
panel.  The Y-type neutral lines extend horizontally along the top
and bottom of the sheet (dotted lines).  The guide field
is directed to the 
right ($+\xhat$) on both sides of the sheet.  The field on the near
side (solid) is directed upward while that on the far side (dashed) is
directed downward.  The angle between these field lines is
$\Delta\theta$.}
	\label{fig:cs_fig}
\end{figure}

The perpendicular extent of the current sheet follows the
two-dimensional model of Greene (\citeyear{Green1965}) and
Syrovatskii (\citeyear{Syrovatskii1971}).  The net current in the sheet, $I$, determines
its width
\be
  \Delta ~=~ 4\sqrt{{|I|/c\over \bpprime}} ~~,
	\label{eq:Delta}
\ee
where $\bpprime>0$ but $I$ may be of either sign.
The sheet extends along the $\pm\yhat$ direction when $I>0$ and along
$\pm\zhat$ when $I<0$, as shown in \fig\ \ref{fig:cs_fig}.
Since $\bpprime$ varies along the separator so does the width of the
current ribbon, as evident in \fig\ \ref{fig:cr}.  
It does, however, carry the same current $I$ over its
entire length, $L$.

The current sheet arises from a hypothesized absence, during pre-flare
build-up, of reconnection required to
achieve a potential field state.  A potential field extrapolated from
the magnetogram of \fig\ \ref{fig:mdi_fp}, contains field lines
linking, for example, the new pole $P04$ to the old pole $N01$.  In
fact roughly 60\% of the flux from $P04$ connects to $N01$ for a total
flux $\Psi^{\rm (v)}_{4\hbox{--}1}=2.8\times10^{21}$ Mx, 
where the superscript is a reminder
that this value pertains to the potential (vacuum) field.  

All new polarities presumably emerged connected only among one
another.  In this case a connection between new and old regions would 
contain zero flux, $\Psi_{4\hbox{--}1}=0$, in contrast to the non-zero
flux in the potential field.  It is this discrepancy
\be
  \Delta\Psi_{4\hbox{--}1} = \Psi_{4\hbox{--}1} 
  -\Psi^{\rm (v)}_{4\hbox{--}1} ~~,
\ee
between the potential field connectivity and the actual connectivity
which leads to the separator current \cite{Longcope2001b}.  For small
values, the net current follows a simple self-inductance relation
\be
  {I\over c}~\sim~ \pm{\Delta\Psi\over 4\pi L} ~~,
	\label{eq:I}
\ee
where we have omitted, for clarity, a factor depending on $\ln(|I|)$.

The sense of current depends on the sense of change, $\Delta\Psi$,
relative to the sense of the potential field.  Potential field links
the separator of \fig\ \ref{fig:cr} in a positive sense: right-handed
following the separator in the direction of the guide field.
Before any reconnection, $\Psi_{4\hbox{--}1}=0$, so the self
flux from the current ribbon must be directed through the separator
in the opposite sense.  Thus the current must flow anti-parallel to
the guide field: $I<0$.  This produces the roughly vertical sheet
shown in \fig\ \ref{fig:cr}.

The separator shown in \fig\ \ref{fig:cr} has length $L=54$ Mm and will
therefore, according to \eq\ (\ref{eq:I}), 
carry $I/c\simeq4\times 10^{10}\,{\rm G\,cm}$ 
(\ie\ 400 GAmps) prior
to any reconnection.  This is confined to a current sheet whose width
at the coronal point ($\bpprime=34\,{\rm G\,Mm^{-1}}$) is
$\Delta\simeq14$ Mm.  It is natural that the half-width of this
vertical sheet is close to the height
of the point since the current sheet must totally separate the new
and old fluxes.  Had the absence of reconnection across this one
separator been the only constraint on the coronal field, the field would
contain free energy
\be
  \Delta W ~\simeq~{I\over 2c}\Delta\Psi ~=~ {\Delta\Psi^2\over 8\pi L} ~~,	
  	\label{eq:DW_val}
\ee
which is $\Delta W=5\times10^{31}$ ergs.
Although there are numerous other separators imposing other constraints, this
crude estimate is consistent with the amount of energy ultimately
released in the flare.

While the width of the sheet, $\Delta$, depends on global properties
of the field, its thickness, $\delta$, depends on non-ideal
processes significant inside the sheet.  
For a Sweet-Parker solution in the presence of uniform
resistivity, $\eta$, the thickness would be 
$\delta\simeq\Delta S_{\Delta}^{-1/2}$ where the current sheet
Lundquist number is
\be
  S_{\Delta} ~=~ {\Delta v_{{\rm A},\perp}\over\eta} ~=~
  {|I|\over I_{\rm sp}} ~~,~~
  {I_{\rm sp}\over c} ~=~ \eta\sqrt{\rho\over 4\pi} ~~.
\ee
Spitzer resistivity in a coronal plasma ($T=10^6$ K and 
$n_{\rm e}=10^{10}\,{\rm cm}^{-3}$) yields a characteristic current
$I_{\rm sp}\simeq10^{-2}$ Amps, for which the Lundquist number will exceed
$10^{12}$.  In this case the equilibrium width of the current sheet
would be $\delta\simeq14$ m.  Sweet-Parker reconnection at such a high
Lundquist number transfers flux very slowly, making it essentially
``non-reconnection''.  In spite of the large current density the
energy dissipation, $P\approx (I/c)d\Psi/dt\approx\Delta W/\tau_{sp}$,
is negligibly small.  Thus we would expect signatures 
of a pre-flare current sheet to be extremely weak, if observable at all.
This is constant with the absence of any current-sheet signature from pre-flare EUV images.

Given this slow reconnection it is possible for the current sheet to
exist for any portion of the 50-hour build-up phase.  
The flare itself must arise from a rapid transfer of flux across
the current sheet.  In other words,
the current sheet is necessary but not sufficient for reconnection.

In light of the coincidental appearance of
numerous thin coronal loops of the new connectivity it seems that
flux transfer was unsteady and spatially intermittent rather than a 
steady, extended electric field along the separator current sheet.  Indeed,
it has become recognized that localized electric fields are a
necessary condition for fast magnetic reconnection
\cite{Biskamp2001,Kulsrud2001}.

\section{A Model of Time-Dependent Localized Reconnection}

The localized reconnection will occur across a current sheet where
field lines of different connectivity come into close enough proximity
to be interconnected by localized processes.  As a model of the
process consider a
Green-Syrovatskii current sheet with a guide field $B_g$ such as that
from \fig\ \ref{fig:cs_fig}.  Such a current sheet
separates field lines of the same
magnitude but differing in angle by
$\Delta\theta=2\tan^{-1}(B_{\perp}/B_g)$.  The strength of the
reconnecting component depends on the width of the current sheet,
$B_{\perp}=\bpprime\Delta/2$, so the reconnection angle is
\be
  \Delta\theta ~=~ 2\tan^{-1}\left({\Delta\over 2B_g/\bpprime}\right)
     ~~.
	\label{eq:Dtheta}
\ee
The separator in our pre-flare field is characterized by a dimension 
$B_g/\bpprime\simeq10$ Mm.  When the current sheet width reaches
$\Delta=14$ Mm the reconnecting field angle is $\Delta\theta=70^{\circ}$.  

\begin{figure}[htp]
\psfig{file=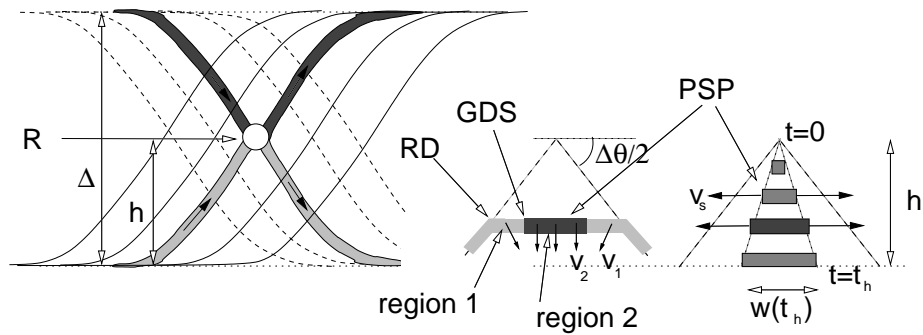,width=4.8in}
\caption{Patchy, transient reconnection in a current sheet.
Reconnection within patch $R$ creates two flux tubes (light and dark
grey) shown in the left panel.  Each of these retracts through the
sheet; the central panel illustrates the retraction of the lower tube.
Rotational discontinuities (RDs)
propagate along the tube at the Alfv\'en speed.  Central compression 
creates two gas dynamic shocks (GDSs) confining a hot, dense
plug of post-shock plasma (PSP).  This moves downward with the flux
tube at speed $v_z$ and expands horizontally at the shock speed
$v_s$.  The right panel shows the plug at successive times culminating in
$t=t_h$, when it reaches the Y-type neutral line along the bottom of
the sheet.}
	\label{fig:patchy}
\end{figure}

The computation above reveals that current sheet reconnection occurs
in the presence of a significant guide field.
A transient and spatially localized reconnection event in such a
current sheet will produce a flux tube as illustrated in \fig\
\ref{fig:patchy}.  While all field lines adjacent to the current sheet
were initially in equilibrium, the 
reconnection produces sudden disequilibrium
from very sharp bends in the new tubes.  These bends produce a rapid
retraction of the tube; each tube pulls itself straight by sliding
through the sheet \cite{Linton2006,Linton2009,LGL09}.  The retraction is governed by dynamics independent of the process which created the
non-equilibrium tubes.

\subsection{Structure of reconnection outflow}

Post-reconnection retraction releases magnetic energy by
shortening the tube on the Alfv\'en time characteristic
of magnetically-driven dynamics.  The initial bend decomposes into two
rotational discontinuities (RDs)
propagating at the Alfv\'en-speed and two
slow mode shocks (SMSs) or gas-dynamic shocks (GDSs).  The RDs produce
the rapid shortening, but they do so without changing temperature or
density  of the plasma.  This rapid shortening takes the form of
Alfv\'en speed motions $\vvec_1$ directed partly inward in order to
decrease the length.  This parallel compression is therefore
much faster than the slow magnetosonic waves which govern its
dynamics, and the flows collide in SMSs or GDSs.

Longcope \etal\ (\citeyear{LGL09}; hereafter called \lgl) 
presented a three-dimensional model of such
transient, patchy magnetic reconnection.  They used thin-flux tube
dynamics to describe the retraction and compression.  They found dynamical evolution following the diagram in the center panel of \fig\ \ref{fig:patchy}, with
purely hydrodynamic GDSs formed by the converging flows.

In the model of \lgl\ each RD in the downward-moving tube
produces a flow, designated region 1, 
directed along the bisector of the bend,
\be
  \vvec_1 ~=~ \mp 2v_{A,0} \sin^2(\Delta\theta/4)\,\xhat
  - v_{A,0} \sin(\Delta\theta/2)\,\zhat
\ee
where $v_{A,0}$ is the Alfv\'en speed in the background magnetic field
outside the current sheet.  The two RDs direct flows inward in
opposing horizontal senses.  In a reference frame moving downward with the
horizontal segment each inflow has gas-dynamic Mach number (\lgl)
\be
  M_1 ~=~ {|v_{x,1}|\over c_{s,1}} ~=~ \sqrt{8\over\gamma\beta_0}
  \sin^2(\Delta\theta/4) ~~,
  	\label{eq:M1}
\ee
where $c_{s,1}$ is the sound speed in the pre-reconnection (and pre-shock) plasma.
Except in cases of very small reconnection 
angles $\Delta\theta$, the small value of pre-reconnection $\beta$, 
denoted $\beta_0$, ensures very large inflow Mach number.

\begin{figure}[htp]
\psfig{file=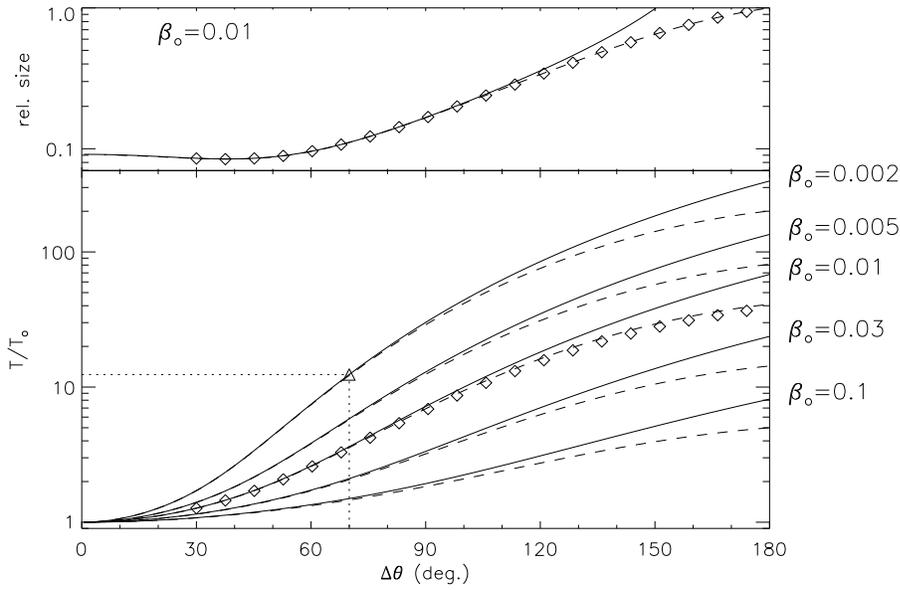,width=4.8in}
\caption{Properties of the PSP {\em vs.} the reconnection angle
$\Delta\theta$ for various models.  Curves shown are
solid: Longcope \etal\ (\lgl), dashed: Soward
(\protect\citeyear{Soward1982b}) and diamonds:
Lin and Lee (\protect\citeyear{Lin1994}).  
The lower panel shows the temperature ratio,
$T_2/T_1$, for various values of the initial plasma-$\beta$, noted on
the right.  All models agree for small angles for each value of
$\beta_0$.  A triangle denotes the situation for a current sheet with $\Delta\theta=70^{\circ}$ and $\beta_0=0.002$.  The top panel shows the fraction of the tube between RDs 
occupied by the PSP for the case of $\beta_0=0.01$ for all three
models.}
	\label{fig:trat}
\end{figure}

The collision of these two inward flows brings the plasma to rest at
two gas-dynamic shocks (GDSs).  The density increase across these
shocks follows from the gas-dynamic conservation laws (\lgl)
\be
  r ~=~ {\rho_{2}\over \rho_{1}} ~=~
  {\sqrt{9 + 4M_1^2} + 2 M_1\over \sqrt{9 + 4M_1^2} - M_1} ~~,
  	\label{eq:denrat}
\ee
assuming the shocks are in steady state (we have used $\gamma=5/3$).  The temperature ratio across the shock
\be
  {T_2\over T_1} ~=~ {1\over r}\left[ \, 1 + 
  \hbox{${5\over3}$} M_1\left(\sqrt{ 1 + 4M_1^2/9} + 2M_1/3 \right)\, \right] ~~,
  	\label{eq:trat}
\ee
depends on $\beta_0$ and $\Delta\theta$ through \eqs\ (\ref{eq:M1}) and 
(\ref{eq:denrat}); this dependence is plotted in \fig\ \ref{fig:trat}.

\subsection{Alternative reconnection models}

In the discussion above the heating follows from non-resistive, post-reconnection dynamics alone.  It is therefore natural that the model of \lgl\ gives results similar to other models of fast magnetic reconnection.  
The most well-known such model, that of Petschek
(\citeyear{Petschek1964}), is of
two-dimensional, steady-state reconnection between perfectly
anti-parallel field: $\Delta\theta=180^{\circ}$.  In that model two long
steady outflow jets are confined between standing SMSs.  It was realized
early on \cite{Petschek1967} that the addition of a guide field
demanded a more complicated shock structure involving 
both RDs and SMSs.  For
$\beta\approx1$, or $\Delta\theta\simeq180^{\circ}$, the shocks are
close together and the configuration appears similar to the
classic Petschek one.  In other cases, such as our own with
$\beta\ll1$ and a significant departure from
$\Delta\theta=180^{\circ}$, the SMSs occur well inside the bends in
the field: the RDs.  From the conventional end-on perspective the
configuration still resembles the Petschek model, but
from the side-view perspective any given field line looks
like the flux tube in the center panel of \fig\ \ref{fig:patchy} but
with SMSs instead of GDSs.

A complete quantitative treatment of
Petschek reconnection in the presence of a guide field was presented by
Soward (\citeyear{Soward1982b}) and then re-derived in simpler form by
Forbes \etal\ (\citeyear{Forbes1989}) and Vrsnak and Skender (\citeyear{Vrsnak2005}).
This analysis reveals that the post-reconnection
temperature depends on angle and Alfv\'en speed alone, in a
manner very similar to the transient three-dimensional \lgl\ model
(see \fig\ \ref{fig:trat}).  It does not, for 
example, depend on the reconnection rate or even
the microphysics generating the reconnection electric field.
The distance along a 
given field line between the SMSs is some fraction of the distance
between the RDs.  In the steady-state, two-dimensional model this
fraction depends on $\Delta\theta$ in exactly the same way as the
post-shock plasma (PSP) size does in the transient three-dimensional model (see top panel
of \fig\ \ref{fig:trat}).

Properties of the more extensively studied Petschek model can be read off the far right side of \fig\ \ref{fig:trat}: $\Delta\theta=180^{\circ}$.  
In that case the two shocks, RD and SMS, combine into the single shock with no gap --- the PSP occupies 100\% of the region, where the dashed curve intersects the right axis of the upper panel.  The final temperature ratio depends on $\beta_0$; the oft-used value $\beta_0=0.01$ produces a 40-fold increase in temperature, in agreement with the dashed curve in the lower panel. 

Figure \ref{fig:trat} reveals remarkably good quantitative agreement, especially 
for for $\Delta\theta\la120^{\circ}$, between seemingly different reconnection 
models.  The level of disagreement at large angles results from
significant reduction in field strength across the SMSs assumed 
absent from the GDSs of \lgl.  This reduction 
drives the post-shock $\beta$ to non-negligible values contradicting
the assumption of \lgl\ that $\beta\ll1$ on both sides.
These assumptions do, however, appear to be good ones for
smaller angles.  In particular, the SMSs do not significantly decrease the magnetic field strength when $\Delta\theta\la120^{\circ}$, in contrast to their switch-off character in the traditional Petschek limit ($\Delta\theta=180^{\circ}$).

The agreement between a steady two-dimensional model and a non-steady
three-dimensional model can be understood in terms of a non-steady
one-dimensional model.  Lin and Lee (\citeyear{Lin1994}) 
solved the Riemann problem
beginning with an infinite current sheet bending an otherwise uniform
magnetic field.  This sheet decomposes into a pair of fast
magnetosonic rarefaction waves, outside a pair of RDs,
outside a pair of SMSs.  It has been demonstrated
that this solution is similar to Petschek reconnection in
two-dimensions either with or without a guide field \cite{Lin1999b}.
The diamonds in \fig\ \ref{fig:trat} show that the SMS separation
(relative to the RD separation) and the post shock
temperature match those of Soward (\citeyear{Soward1982b}) 
over all angles.   Even a steady, magnetized flow contains individual flux tubes each undergoing time-dependent evolution.   This evolution obeys non-resistive MHD, independent of the reconnection mechanism.  Evidently it is equally well described in one, two or three dimensions.

We conclude that post-shock temperatures from fast
reconnection across a current sheet depends on the field strength and
angle according to a fairly robust relationship given by \eq\ (\ref{eq:trat}) and
shown in \fig\
\ref{fig:trat}.  The specific details of the reconnection appear to be
irrelevant in determining this particular property.  The reconnection 
outflow may take the form of a long jet, as many models have assumed, or
it may consist of numerous isolated plugs.  Either situation produce very similar post-shock properties.  For the initial angle
$\Delta\theta=70^{\circ}$ to heat a plasma from active region
temperature, $T_0=3$ MK, to $T_2=37$ MK observed in the loop-top
source, requires reconnection at $\beta_0\simeq0.002$.  In a field
strength of $B=B_g/\cos(35^{\circ})=400$ G, a 3 MK has this value of beta for
$n_{\rm e,0}=1.5\times10^{10}\,{\rm cm}^{-3}$.

\section{A Model of the Loop-top Source}

We suggest that the loop-top thermal emission observed by RHESSI is from a
post-shock plasma of the kind just described.  The numerous distinct loops 
observed in 171\AA\ suggest a spatially intermittent (patchy) 
reconnection has formed isolated high-temperature plugs of the kind shown in \fig\
\ref{fig:patchy}.  The flux tubes would be formed by reconnection
across a current sheet like that in \fig\ \ref{fig:cr}.  Each plug 
has the same cross-sectional area as the
tube and extends a distance $w(t)$ along its axis, as in
\fig\ \ref{fig:patchy}.   With flux
$\delta\varphi$ and field strength, $B$, set by the flux layers
between which it is confined, the flux tube has cross sectional area
$\delta\varphi/B$.  This is independent of the actual thickness of the current sheet across which the reconnection occurs; in fact the \lgl\ model and the simulation which motivated it \cite{Linton2006,Linton2009} take the sheet to much thinner than the tube.  The emission measure of this one plug is
\be
  \delta EM ~=~ n_{\rm e,2}^2\,w\,\delta\varphi/B ~~,
\ee
where $n_{\rm e,2}$ is the post-shock electron density.  Provided the
plasma is sufficiently dense to thermalize, all of
the plug's emission is characterized by the post-shock temperature $T_2$.

New patches of reconnection, distributed over the current sheet,  will
produce new flux tubes, two per patch, 
at an average rate $2\dot{\Phi}/\delta\varphi$.
Each of these includes a central plug which emits for a time $\tlife$.
Emitting plugs have a combined emission measure
\be
  EM ~=~ 2n_{\rm e,2}^2\,\avg{w}\,\tlife{\dot{\Phi}\over B_0} ~~.
	\label{eq:EM}
\ee
where $\avg{w}$ is the axial extent averaged over the
emitting life-time.  The emission measure therefore depends, through 
properties on the right hand side, on the angle $\Delta\theta$ between
reconnecting field lines and the mean reconnection rate.  This rate,
$\dot{\Phi}$ is not related to the reconnection electric field within
a single patch, but is rather the rate at
which new patches are produced throughout the current sheet.

\subsection{The size and life of the PSP plug}

The post-shock plasma (PSP) plug is confined by supersonic inflows
generated by the rotational discontinuities (RDs) shown in \fig\
\ref{fig:patchy}.  The size, $\avg{w}$ and life, $\tlife$,
of a single emitting plug can be estimated by considering how long these
flows are able to confine it.  This follows from a detailed analysis of the
plasma dynamics inside the retracting flux tube shown in \fig\
\ref{fig:patchy}.

We hereafter assume reconnection at sufficiently low $\beta_0$ 
(pre-reconnection) and at sufficiently large angle, $\Delta\theta$, that the flow in region 1 is hypersonic:  $M_1\gg1$ according to \eq\ (\ref{eq:M1}). 
Using this in  \eq\ (\ref{eq:denrat}) gives
density ratio near the gas-dynamic limit $r\simeq4$.  We hereafter use this
value to simplify the analysis; using the full expression would produce more complex expressions but similar values.
The accretion of the inward flows cause the PSP to grow as the GDSs
propagate at 
\be
  v_s ~=~ {v_{x,1}\over r-1} ~\simeq~ \third v_{x,1} ~~.
\ee

Vertical motion of the tube segment persists at roughly
constant speed, $v_{z,1}=-v_{A,0}\sin(\Delta\theta/2)$, until it
reaches the bottom of the current sheet.  This happens 
at $t_h=h/|v_{z,1}|$ (see \fig\ \ref{fig:patchy}), at which time
the PSP occupies
\be
  w(t_h) ~=~ 2v_s\,t_h ~=~ \hbox{${2\over3}$}h\tan(\Delta\theta/4) ~~,
\ee
of the total segment length, $2h\cot(\Delta\theta/2)$.  This is the fraction whose dependence on $\Delta\theta$ is plotted in \fig\ \ref{fig:trat}.  In purely two-dimensional models, \ie\ $\Delta\theta=180^{\circ}$, the retraction is halted at a fast magnetosonic termination shock \cite{Forbes1983}; here it is a single event in the evolution of the tube and the PSP.

\begin{figure}[htp]
\psfig{file=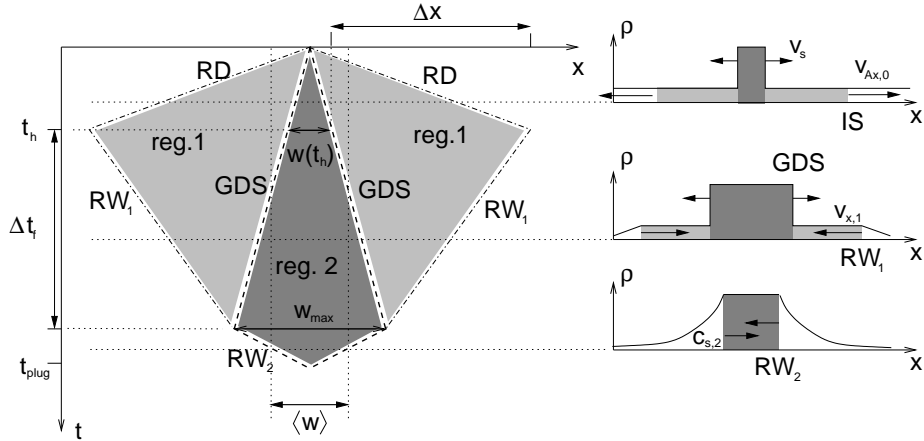,width=4.8in}
\caption{A schematic of the time evolution of the post-shock plasma
plug depicted with time running downward.  The three graphs on the
right show density {\em vs.} $x$ at characteristic times.
Regions 1 and 2 are light and dark grey regions confined by shocks and
rarefaction waves moving at different horizontal velocities.  The
panel on the left shows the evolution of these regions and the
different shocks.  Horizontal arrows show the extent of the PSP plug
at specific times.}
	\label{fig:psp}
\end{figure}

The end of the RDs means the end of the supersonic inflows confining
the PSP.  The vertical forces which halt the retraction do
not, however, directly stop the horizontal flow throughout region 1.
Instead, inflow stoppage propagates horizontally as a
rarefaction wave (RW$_1$ in \fig\ \ref{fig:psp})
moving at $v_f=c_{s,1}+v_{x,1}\simeq v_{x,1}$.
At time $t_h$ the rarefaction wave and the shock are separated by a
horizontal distance
\be
  \Delta x ~=~ h\cot(\Delta\theta/2)- \half w(t_h) 
  ~=~ h[\cot(\Delta\theta/2) - \third \tan(\Delta\theta/4) ]~~,
\ee
as shown in \fig\ \ref{fig:psp}.
They move toward one another at relative speed,
$v_f+v_s$, meeting after 
\be
  \Delta t_f ~=~ {\Delta x\over v_{x,1}+v_s} ~=~ {h\over 4v_{x,1}}\left[
  3\cot(\Delta\theta/2) - \tan(\Delta\theta/4)\right] ~~.
\ee
At this time the shock has propagated a horizontal distance
\be
  w_{\rm max} ~=~2 v_s( t_h + \Delta t_f) ~=~ 
  {h\over 2\sin(\Delta\theta/2)}
\ee
since the reconnection episode; this is the maximum extent of the plug
(see \fig\ \ref{fig:psp}).

The interaction of the rarefaction wave and the gas-dynamic shock
results in a new rarefaction wave (RW$_2$) 
propagating toward the center of the
plug at the post-shock sound speed 
$c_{s,2}$.  This represents the free expansion of the unconfined hot
plasma which ultimately disassembles the plug.  These second rarefaction waves will meet after an interval
\be
  \Delta t_c ~=~{w_{\rm max}\over 2 c_{s,2}} ~=~
  {w_{\rm max}\over 2 v_{x,1}}M_1\,\sqrt{{T_1\over T_2}} ~=~
  {w_{\rm max} \over v_{x,1}}{3\over 2 \sqrt{5}} ~~,
	\label{eq:Dtc}
\ee
marking the end of the PSP.
The final expression in \eq\ (\ref{eq:Dtc}) uses the high Mach-number
limit  of \eq\ (\ref{eq:trat}), $T_2/T_1\to(5/9)M_1^2$. 
The total lifetime of the plug
\be
  \tlife ~=~ \half w_{\rm max}
  \left[{1\over v_s} + {1\over c_{s,2}}\right]
  ~=~ {3w_{\rm max}\over 4v_{A,0}\sin^2(\Delta\theta/4)}\left[1 + 
  {1\over\sqrt{5}}\right] ~~,
	\label{eq:tlife}
\ee
is longer than its free expansion by a factor $c_{s,2}/v_s\simeq\sqrt{5}$.
Since the plug expands linearly and then contracts linearly its
average extent is one-half its maximum:
\be
  \avg{w} ~=~\half w_{\rm max} ~=~ {h\over 4\sin(\Delta\theta/2)} 
\ee

In the analysis above the life time of the emitting plasma was fixed
by the time-scale for pressure confinement.  Several previous
investigations have used instead a time-scale for
dissipation by conductive cooling
\cite{Jiang2006}.  A detailed analysis by Guidoni and Longcope (2010)
\nocite{Guidoni2010} shows that conductivity produces extended thermal
fronts in advance of the GDSs.  Temperature increases gradually 
over the fronts reaching the steady-state Rankine-Hugoniot value at
the density jump which is the actual GDS \cite[more accurately called an
iso-thermal sub-shock]{Kennel1987,Xu1992,Guidoni2010}.  There is
very little density enhancement within the fronts, so they would not
contribute appreciably to the emission measure.  Thermal conduction
could precede the rarefaction wave RW$_2$, causing $\Delta t_c$ to be
even smaller than our estimate in \eq\ (\ref{eq:Dtc}).  
This is the smallest contribution to
$\tlife$ so we expect that 
\eq\ (\ref{eq:tlife}), based on pressure confinement alone, is probably
close to the actual life.

\subsection{The loop-top emission}

Assuming the reconnection occurs near the center line of the
current sheet, of width $\Delta$, the net retraction distance is 
$h=\Delta/2$.  The current sheet width is related to the
reconnection angle by \eq\ (\ref{eq:Dtheta}), 
\be
  h ~=~ {B_g\over\bpprime}\tan(\Delta\theta/2) ~~,
	\label{eq:h}
\ee
from the structure of the Green-Syrovatskii sheet with a guide field.
The field properties at the separator point in \fig\ \ref{fig:cr}
and an angle $\Delta\theta=70^{\circ}$ inferred from RHESSI
temperatures gives an average plug size of $\avg{w}=3.0$ Mm.  Taking
the external density to be $n_{\rm e,0}=1.5\times10^{10}\,{\rm cm}^{-3}$ gives a
life-time of $\tlife=8.5$ sec.  The loop-top source is
somewhat larger and considerably longer-lived than a single PSP so it
must consist of many such events at any one time.

The net emission measure from the assembly of plugs is given by 
\eq\ (\ref{eq:EM}).  Expressing the total magnetic field in terms of
the guide field, $B_0=B_g/\cos(\Delta\theta/2)$, and using \eq\
(\ref{eq:h}), gives
\be
  EM ~=~ 15.4{ n_{\rm e,0}^{5/2}\sqrt{m_p}\over B_{\perp}^{\prime2}}\,
  \,{\dot{\Phi} \over \sin^2(\Delta\theta/4)}~~,
	\label{eq:EM_final}
\ee
where $m_p=\rho/n_{\rm e}$ is the proton mass.  In this way the emission
measure in super-hot thermal material is directly proportional the rate
of reconnection.

There are several ways by which we might derive the reconnection rate
$\dot{\Phi}$.  Equations (\ref{eq:Delta}), (\ref{eq:I}) and
(\ref{eq:Dtheta}), concerning the Green-Syovatskii current sheet, can
be combined to yield a relation between flux transfer and the decrease
in reconnection angle
\be
  \dot{\Phi} ~=~ -{d|\Delta\Psi|\over dt} ~=~
  -\pi L {B_g^2\over\bpprime}\,{\sin(\Delta\theta/2)\over
  \cos^3(\Delta\theta/2)}\,{d(\Delta\theta)\over dt} ~~.
	\label{eq:zeta_volt}
\ee
The reconnection angle $\Delta\theta$ decreases as the current
decreases due to reconnective flux transfer across the
sheet.  The temperature of the PSP also depends on $\Delta\theta$ as
shown in \fig\ \ref{fig:trat}.  It is the decrease in this angle,
due to reconnection, which leads to the decreasing temperature in the
super-hot thermal component shown in \fig\ \ref{fig:RHESSI_T}.  We can
combine these two relations to derive $\Delta\theta(t)$ from $T(t)$
for $B_g=331$G and an assumed value of pre-reconnection density,
$n_{\rm e,0}$.

\begin{figure}[htp]
\psfig{file=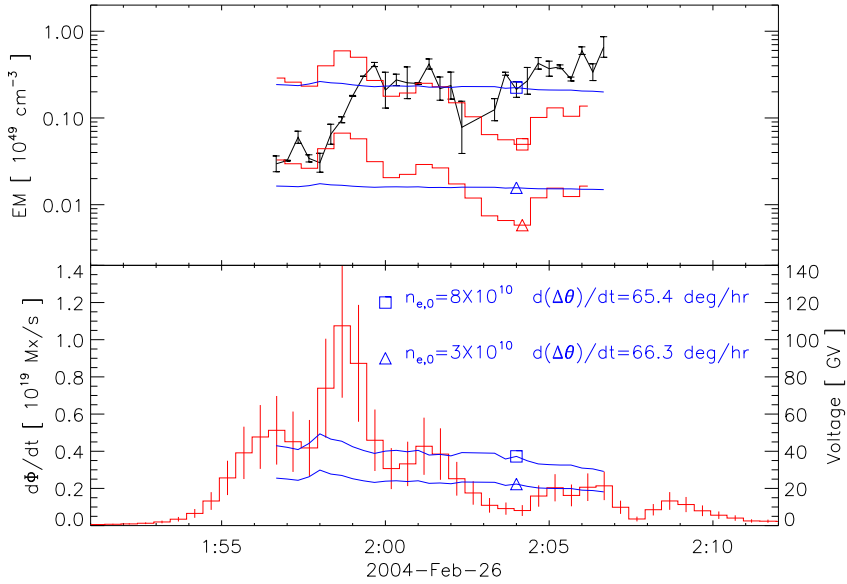,width=4.8in}
\caption{Plots of inferred flux transfer rates (bottom) and the
emission measure they would produce.  Blue curves are derived from the
temperature change using \eq\ (\ref{eq:zeta_volt}) and red curves are
from the motion of 171\AA\ ribbon features across the magnetogram.
Curves with triangles (squares) are derived using 
$n_{e,0}=3\times10^{10}\,{\rm cm}^{-3}$ 
($n_{e,0}=8\times10^{10}\,{\rm cm}^{-3}$).  The black 
curves are the super-hot emission measure derived from RHESSI, error bars are the values from detectors 3 and 4 alone.}
	\label{fig:phidot}
\end{figure}

Assuming larger densities $n_{\rm e,0}=3$ -- $8\times 10^{10}$ results in
slightly larger reconnection angles
$\Delta\theta=80^{\circ}$ -- $100^{\circ}$ due to larger $\beta_0$.
The decreasing temperature causes the angle to decrease at 
$-\Delta\theta/dt\simeq65$ degrees per
hour for the two density choices.  Using these rates and angles in \eq\
(\ref{eq:zeta_volt}) leads to $\dot{\Phi}\simeq20$ -- 50 Gigavolts as
shown in \fig\ \ref{fig:phidot}.  This flux transfer rate in \eq\
(\ref{eq:EM_final}) gives an emission measure commensurate with RHESSI
observations when the larger initial density, 
$n_{\rm e,0}=8\times10^{10}\,{\rm cm}^{-3}$ is used.

A more direct,
and more conventional, measurement of $\dot{\Phi}$ can be obtained by
mapping the apparent motion of flare ribbons across the
magnetic field in the lower atmosphere
\cite{Forbes1984,Poletto1986}.  This is frequently done with high
cadence TRACE UV (1600\AA) images \cite{Saba2006,Longcope2007}, but
these were unavailable at this time since TRACE was imaging instead
in EUV 171 \AA. Observations at this wavelength usually reflect $\ge $1 MK
plasmas in the corona, however, the bandpass does include contribution by enhanced chromospheric emission at the feet of flare loops
especially in the early phase of the flare.  Such emission is evident in the 1:59:14 and 2:03:14 panels of \fig\ \ref{fig:tr_sum}.

We therefore use 171\AA\ images to compute the reconnection flux transfer rate.  We outline the chromospheric ribbons seen at the start of the flare, and, as the flare evolves, track only the outward spread of the ribbon, \ie\ the spread of the brightening away from the polarity inversion line. This approach avoids contamination  due to emission by multi-megakelvin plasmas in the post-flare loops, which have cooled down to emit in 171\AA\ between the expanding ribbons. 

The flare ribbons are tracked in co-aligned TRACE 171~\AA\ images from 1:00~UT for two hours with a cadence
of approximately 40~s and a pixel scale of 0.5~\arcsec. The chromospheric ribbons are then mapped in a magnetogram co-aligned with the flare images and the total magnetic flux covered by the ribbons is measured. To approximate the magnetic field in the chromosphere, we extrapolate
the photospheric line-of-sight magnetic field, obtained by MDI/SoHO 20 minutes before the flare onset, into a nominal chromosphere height of 2000 km using a potential field assumption. Numerous experiments have shown that measurement of the reconnection flux is quite robust \cite{Qiu2007}. The extrapolation merely decreases the measured reconnection flux by 25\%, but does not modify the time profile of $\dot{\Phi}$. Assumptions other than a potential field assumption (such as a linear-force-free assumption) and/or extrapolation to a different height $\le$ 2000 km would not change the
$\dot{\Phi}$ profile, and its magnitude would be modified by $\le$25\%.

We also estimate uncertainties in the flux measurement by mis-aligning MDI magnetogram and TRACE images by up to 2\arcsec\ and by using a set of brightness threshold values for ribbon detection. The upper-limit of the uncertainty due to these sources is 30\%. Given the geometry of this flare, the flare ribbon in the negative magnetic field is
more accurately tracked. Therefore $\dot{\Phi}$ measured in the negative field is used and plotted in red on \fig\ \ref{fig:phidot}.  Using this rate in \eq\ (\ref{eq:zeta_volt}) gives independent measures of the super-hot emission measure.  Once again we find agreement when the larger initial density, $n_{e,0}=8\times10^{10}\,{\rm cm}^{-3}$ is used.

Relation (\ref{eq:zeta_volt}) could be inverted to use the observed
super-hot emission measure to estimate $\dot{\Phi}$.  Agreement in
\fig\ \ref{fig:phidot} shows that doing so yields a third independent
measure in agreement with the other two -- temperature decrease and
ribbon motion.  It appears that the flare is driven by flux transfer
at a mean rate $\dot{\Phi}\approx4\times10^{18}$ Mx/s (40 Gigavolts).
We stress once more that this is not a local electric field, but rather the net effect of multiple reconnection events.

\section{A Model of the flare}

\subsection{The loop-top source}

\begin{figure}[htp]
\centerline{\psfig{file=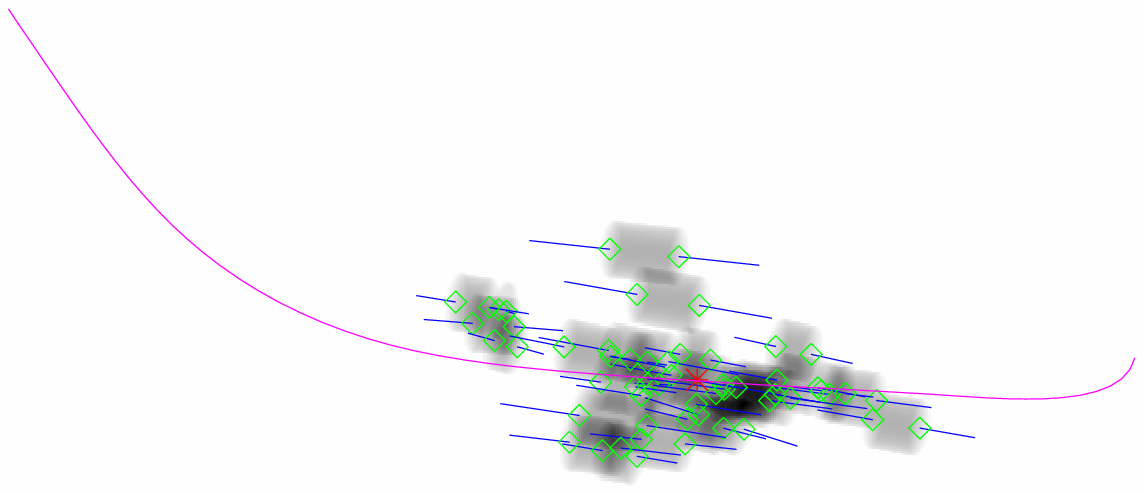,width=3.5in}}
\centerline{\psfig{file=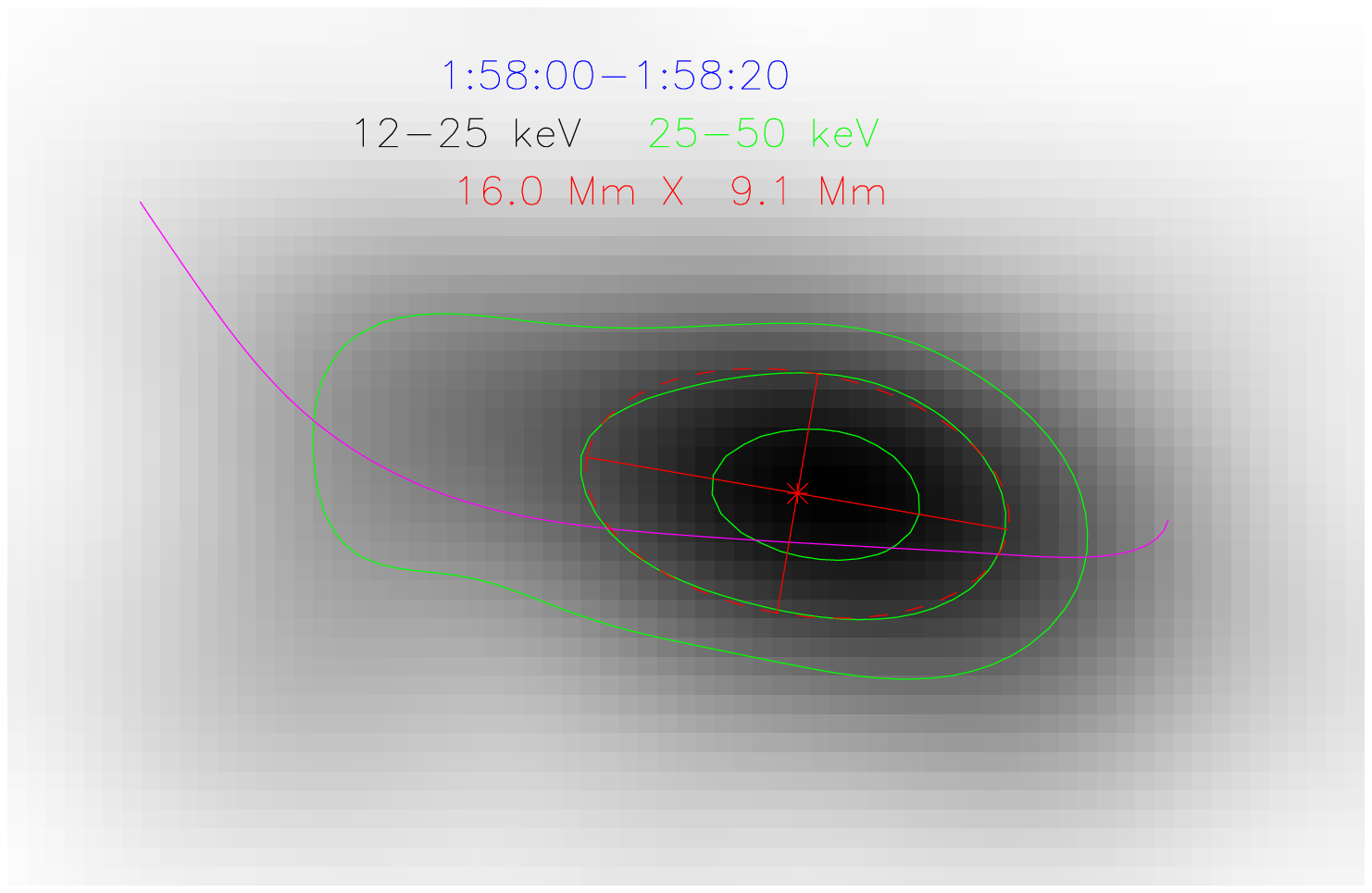,width=3.5in}}
\caption{Images of the loop-top source.  (top) A super-position of 28 plugs 2 Mm across with lengths varying randomly to 6 Mm.  They are distributed randomly in space near the separator (magenta curve).   Blue lines show the extent of the RD along the field lines, and green diamonds show the locations of the GDSs.  A red asterisk marks the centoid of the super-positon.  (bottom) RHESSI images from 12 -- 25 kev (inverse grey scale) and 25-50 keV (green contours) from an integral over 1:58:00 -- 1:58:20, at the peak of the thermal phase.  The red curves show an ellipse with moments matching the central (70\%) contour of the 25-50 keV emission.  The major and minor axes, 16.0 Mm and 9.1 Mm, are straight lines.}
	\label{fig:synth}
\end{figure}

The TRACE loops appearing in the 171\AA\ images (i.e\ \fig\ \ref{fig:tr_sum}) 
have typical diameters of
2 Mm.  Attributing to them the same field strength as the separator, 
$B\approx 330$ G, means each has flux $\delta\varphi=10^{19}$ Mx.
Patchy reconnection at the inferred rate,  $\dot{\Phi}=4\times10^{18}$ Mx sec$^{-1}$, 
would produce such flux tubes
at the rate of $2\dot{\Phi}/\delta\varphi=1$ per second.  The PSP plug
at the center of each tube would persist for $\tlife\simeq8$ sec, so
there would be $\approx8$ such plugs visible at any time.  Using the 
$\Delta\theta=100^{\circ}$ angle inferred from the emission measure computations means the plugs are confined between $|\bvec|=430$ G field, and are thus
$1.7$ Mm in  diameter; they range in length from zero to $w_{\rm max}=5$ Mm.  
We posit that the $9\times16$ Mm 
loop-top source in the 25 -- 50 keV  RHESSI images 
(\fig\ \ref{fig:rh_img}b -- c) is such a superposition of plugs.  
While 8 plugs are visible at any instant, each image shows
a super-position of $\approx28$ plugs 
visible for some portion of its 20-second integration (see \fig\ \ref{fig:synth}).

The volume of a super-hot plug would average 
$\avg{w}\delta\phi/B\simeq5\times10^{24}\,{\rm cm}^{3}$, so the combined 
volumes of the 8 plugs simultaneously visible at one instant is
$V=4\times10^{25}\,{\rm cm}^{3}$. Assigning to the observed 
loop-top source a volume
$V_{\rm s}=\pi(4.5\,{\rm Mm})^2(16\,{\rm Mm})=10^{27}\,{\rm cm}^{3}$, as estimated by
the $70\%$ contour of 25 -- 50 keV images in \fig\ \ref{fig:synth}, it would have a filling factor of $f=0.04$.  (Had we used the 50\% contour instead, the filling factor would have been $f=0.01$.)  A post-shock density of $n_{\rm e,2}=4n_{\rm e,1}=3\times 10^{11}\,{\rm cm}^{-3}$ gives the ensemble of plugs an emission measure of $EM=4\times10^{48}\,{\rm cm}^{-3}$, in good agreement with \figs\ \ref{fig:RHESSI_T} or \ref{fig:phidot}.  

Transferring the entire domain flux,
$\Delta\Psi_{4\hbox{--}1}=2.8\times10^{21}$ Mx, would 
take $\tau_{\rm rx}\simeq12$ minutes at a steady reconnection rate 
$\dot{\Phi}=4\times10^{18}$ Mx sec$^{-1}$ inferred above.  
Each reconnection event transfers flux $\delta\phi=10^{19}$ creating two post-reconnection loops --- one in each post-reconnection domain.  Thus we expect 
$\approx560$ flux tubes will be created over $\approx 700$ seconds (12 minutes).  It is this transfer rate, rather than cooling following impulsive heating, that sets the overall flare time scale.

\subsection{Post-flare loops}

We have presented a model for how fast reconnection could produce super-hot
plasma just outside the diffusion region.  One plug of this super-hot plasma 
would persist for $\tlife\simeq8$ seconds before the confining flows cease and
its own high pressure caused its 
disassembly.  Thermal conduction during and following the plug's
compression would eventually reach the footpoints of the flux tube
causing evaporation.  Details of this phase are beyond the scope of
our model, however, it is possible to predict its overall contribution.

We suggest that the evaporation from the thermal conduction fills the loop with 
lower-temperature plasma observed as the cooler ($T\simeq20$ MK) 
component of the RHESSI spectra.  Following Antiochos and Sturrock 
(\citeyear{Antiochos1978}) we assume the conductive cooling transfers all energy from the thermal and kinetic energy of the tube to the evaporated component.  The temperature and emission measure of this component matches the plasma observed in X-rays by GOES 
(see \fig\ \ref{fig:RHESSI_T}).  Such observations are commonly explained in terms of the Neupert effect \cite{Neupert1968,Dennis1993} whereby direct flare energy release evaporates chromospheric material.  We propose here that, at least in this present case, 
the evaporation is driven by thermal conduction rather than non-thermal particles.

Images from RHESSI's  6 -- 12 keV often resemble other soft-ray 
images and can be used to estimate the extent of the plasma detected by GOES.  The 50\% red contours \figs\ \ref{fig:rh_img}a -- d outlines a $30\times16$ Mm region, suggesting an ellipsoid of volume $V=4\times10^{27}\,{\rm cm}^3$.  Using a filling factor, $f=0.18$, justified below, we estimate the density of evaporated
material as $n_{\rm e}=\sqrt{EM/f V}$.  Using $EM$ from ratio of 
GOES channels gives a maximum
$n_{\rm e}=2.6\times10^{11}\,{\rm cm}^{-3}$ at 2:04:12, thereafter
falling steadily to $n_{\rm e}=6\times10^{10}$ at 3:00:00.

Once a fully retracted tube is full of $T\approx20$ MK plasma from evaporation 
it will cool through thermal conduction and then radiation.   According to the cooling model of Cargill \etal\ (\citeyear{Cargill1995}), a loop at the temperature and density inferred from GOES, and length given by \eq\ (\ref{eq:Lloop}), would cool to half its temperature in between 1 minute (at 1:58) and 3 minutes (at 2:30).  They would reach $T\simeq 1$ MK shortly thereafter and appear in the TRACE image.  This explains loops visible by TRACE almost immediately, even as the temperature inferred by GOES approaches 1 MK much more gradually.  The continued appearance of the these loops over one hour is due to the continued creation of new loops through reconnection.  

\subsection{Energetics}

 Each tube of reconnected  flux $\delta\phi=10^{19}$ Mx releases energy by decreasing its length by a distance $\Delta\ell=2h\tan(\Delta\theta/4)$.  This decrease in length, with no appreciable change in field strength, will decrease the flux tube's magnetic energy by
\be
  \delta E ~=~ {B\delta\phi\over8\pi}\Delta\ell ~\simeq~ 
  10^{29}\,{\rm erg} ~~,
\ee
using values inferred for the current sheet, with $h=7$ Mm and 
$\Delta\theta=100^{\circ}$, $B=430$ G.  The expansion of the external flux layers into the volume vacated by the retraction will release an equal amount of energy (\lgl) leading to a 
net release of $\delta E_{\rm mag}=2\times10^{29}$ ergs.  This is the value at the onset of the flare (1:58:00); later flux tubes will release steadily diminishing energies as the current sheet shrinks causing $\Delta\theta$ and $h$ to decrease.   If the average energy release is half the maximum value, the complete set of 560 flux tubes would release $5.6\times10^{31}$ ergs, consistent will the initial free energy, $\Delta W$, from \eq\ (\ref{eq:DW_val}).

At a time $t_h=3$ sec following its reconnection a given tube has fully retracted and released all its magnetic energy.  At this instant the plug is $w(t_h)=2.2$ Mm long and contains
thermal energy
\be
  U^{\rm (plug)}_{\rm th}(t_h)=3k_{\rm b} T_2 n_{e,2}(\delta\phi/B)\,w(t_h) 
  ~\simeq~2\times10^{28}\,{\rm ergs}
\ee
only $10\%$ of  the magnetic energy liberated by the reconnection,
$\delta E_{\rm mag}$.  This is rather low efficiency compared to the anti-parallel 
case ($\Delta\theta=180^{\circ}$) where $40\%$ of the free energy is directly thermalized by the slow shocks \cite{Priest2000}.  In our case
the other $90\%$ of the liberated energy was converted 
into bulk kinetic energy of retraction.  Roughly  one-sixth ($\sin^2[\Delta\theta/4]$) of that remaining kinetic energy is directed horizontally and is ultimately thermalized before the plug disassembles itself.  Thus $25\%$ of all magnetic energy liberated by reconnection appears as thermal energy in the loop-top source at some point.  Since a given plug will have, on average, half of this amount at a random time, the eight plugs simultaneously visible will contain 
\be
  U^{\rm (sh)}_{\rm th} ~=~ 8\times 3k_{\rm b} T_2 n_{e,2}(\delta\phi/B)\,\avg{w} 
  ~\simeq~ 10^{29}\, {\rm erg} ~~.
  	\label{eq:Uth}
\ee
This is our estimate of the thermal energy contained in the super-hot loop-top source at its peak.

\begin{figure}[htp]
\centerline{\psfig{file=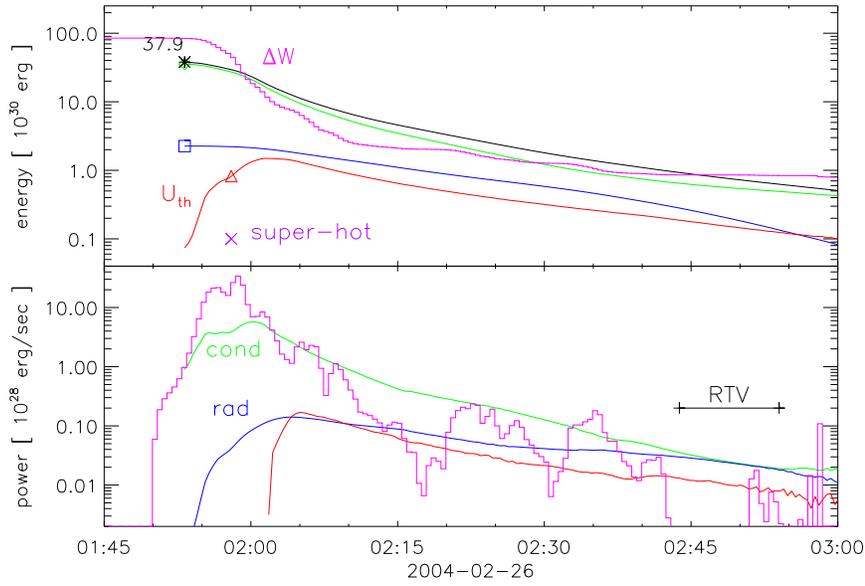,width=4.8in}}
\caption{Time histories of the energy (top) and power (bottom) from the flare.  Green and blue curves are losses from conductivity and radiation respectively.  These losses are integrated backward to produce cumulative energies in the top panel --- the net energy lost {\em after} that time.  The red line in the top panel is the thermal energy content,
$U^{\rm (sh)}_{\rm th}$ from \eq\ (\ref{eq:Uth}), in the lower panel it is 
$-dU^{\rm (sh)}_{\rm th}/dt$.  The instantaneous thermal energy content of the 
low-temperature and high-temperature components at 1:58:00 
are shown in the top panel by $\diamond$ and $\times$ respectively.}
	\label{fig:net_erg}
\end{figure}

The instantaneous thermal energy content of the low-temperature component, filled with evaporated material, is $U_{\rm th}^{\rm (GOES)}=3k_{\rm b}T\sqrt{EM\, Vf}$, 
where we take $T$ and $EM$ from the GOES channels and assume a 
filling factor $f=0.18$ for reasons described below.  The thermal energy, shown in 
\fig\ \ref{fig:net_erg}, rises to a peak of $2\times10^{30}$ ergs 
at 2:01:25.  At 1:58:00, when we found the super-hot component energy, the 
low-temperature component has an order of magnitude more (red triangle).

The low-temperature component cools by radiation and by thermal conduction.  Radiative losses, computed using the temperature dependent Mewe radiative loss function 
\cite{Mewe1985}, are plotted in blue on \fig\ \ref{fig:net_erg}.  
The rate of conductive energy loss to the chromosphere is 
$P_{\rm cond}=U_{\rm th}/\tau_{\rm cond}$ where 
\be
  {1\over \tau_{\rm cond}} ~=~{\kappa|\nabla T|\over 3k_{\rm b}Tn_{\rm e}(L/2)}
  ~\simeq~ {\kappa_0T^{5/2}\over 3k_{\rm b}n_{\rm e}(L/2)^2} ~~,
\ee
for a loop of full length $L$;
Spitzer conductivity is $\kappa=\kappa_0T^{5/2}$.  The EUV loops 
traced in \fig\ \ref{fig:loops} had apparent (plane-of-the-sky) lengths increasing from 
7 Mm to 12 Mm according to empirical relation \eq\ (\ref{eq:Lloop}).  Doubling this time-dependent function to account, approximately, for the third dimensions 
and portions not observed at 171 \AA\, 
gives conductive cooling times ranging from $\tau_{\rm cond}=10$ sec at 2:00, to 
$\tau_{\rm cond}=280$ sec at 3:00.  The resulting conductive losses are plotted in 
\fig\ \ref{fig:net_erg}.  These losses actually represent a transfer of energy to the lower temperature atmosphere, where it is presumably radiated.  This radiative loss is known to far exceed the radiation from optically thin coronal plasma inferred from the 
GOES X-ray bands \cite{Emslie2005}.  We use the conductive losses to approximately quantify this other loss mechanism.

At $T\simeq20$ MK conductive losses dominate radiative losses by up to an order of magnitude.  We chose filling factor $f=0.18$ in order that the two become equal at late times, designated {\sf RTV} in \fig\ \ref{fig:net_erg}, as they should be 
in a static equilibrium \cite{Rosner1978}.  Prior to that, 
short conductive cooling times, 10 -- 280 seconds, demand a sequence of new, 
dynamic loops \cite{Warren2002,Warren2006}.

Integrating each of the losses gives the cumulative losses shown in the top panel 
of \fig\ \ref{fig:net_erg}.  The total integrals of radiation and conduction, 
$2.3\times10^{30}$ ergs and $3.3\times10^{31}$ ergs, are shown by a square 
and diamond, respectively.  We estimate the total energy lost by the low energy 
component as the sum of these two $\Delta W=3.8\times10^{31}$ (asterisk).  
This is a slightly smaller than the MCC estimate of free magnetic energy, 
\eq\ (\ref{eq:DW_val}), suggesting, among other possibilities, incomplete magnetic 
relaxation.

To ascertain the degree of reconnection we integrate the ribbon-swept flux whose 
change, $\dot{\Phi}$ is plotted \fig\ \ref{fig:phidot}.  The flux discrepancy
\be
  \Delta\Psi(t) ~=~ \int_t^{\infty}\dot{\Phi}(t')\, dt' ~~,
\ee
is the same as the ribbon flux, but accumulated from the final time.  The initial value of this flux, $\Delta\Psi=3.3\times10^{21}$Mx, is 20\% larger than $\Delta\Psi_{\hbox{4--1}}$.  
The later is only a single post-reconnection domain, albiet the largest one, so it is perhaps unsurprising that it falls short of the total flux swept by the flare ribbons.  Using the ribbon-swept flux in the magnetic energy estimate, \eq\ (\ref{eq:DW_val}), with the same separator $L=54$ Mm, gives an estimate of the magnetic energy liberated by reconnection.  The total, 
$\Delta W=8.5\times 10^{31}$ ergs, half-again as large as the estimate using 
$\Delta\Psi_{\hbox{4--1}}$ and twice the energy 
released by the flare.  Indeed, throughout the thermal phase of the flare, after 1:56:40, 
the free magnetic energy tracks the total flare energy very well.  It seems that energy release 
through magnetic reconnection is a viable candidate for producing the observed 
flare emission.

\section{Discussion}

We have presented here a new model for the shock heating of plasma as magnetic energy is released following magnetic reconnection.  The flux transfer achieved by the reconnection electric field {\em triggers} this energy release but does not directly participate in it.  As a result its details do not enter the prediction, and we have bypassed their consideration.  The model presented here applies a novel treatment of transient, localized  reconnection in three dimensions, but its shock heating is closely related to the traditional two-dimensional, steady-state model of Petschek.  The role of shocks are similar in both cases, but the former does not occur in an out-flow jet geometry.

We have applied this model to RHESSI, TRACE, MDI and GOES observations of a particular compact X-class flare.  Magnetic energy is stored during a 50-hour emergence of flux within the active region, and then released by reconnection over less than one hour.  The majority of energy released in the flare appears to occur through the production of a super-hot ($T\ga30$ MK) thermal loop-top source, rather than through the acceleration of non-thermal particles.  Our model predicts temperature and emission measure of the source, roughly consistent with those inferred from fitting the RHESSI spectra.  

The observed source is modeled as a super-position of short-lived plugs of plasma heated and compressed by the post-reconnection shocks.  While this ensemble never contains more than $1\%$ of the magnetic energy ultimately released by the flare, it is the first stage in the entire energy release process.  The subsequent expansion and conductive cooling of the plugs drives chromospheric evaporation to produce the more plentiful low-temperature plasma ($T\la20$ MK) observed in soft X-rays.  These cool freely to appear minutes later, when $T\simeq1$ MK, in TRACE EUV images.  The numerous discrete loops are a direct consequence of the transient, localized reconnection which creates them; their early appearance is a result of the rapid, independent evolution of each loop.  The smooth evolution of the ensemble is a result of the gradual decrease in current as flux is transferred.

We assumed the reconnection occurred within a current sheet which had formed at some time prior to the flare.  Current sheets are pre-requisites for traditional fast reconnection models since they bring field lines of significantly different connectivity close enough to be mutually affected by a small-scale reconnection electric field. We estimated the properties of the current sheet assuming it be in equilibrium or reconnecting slowly according to the Sweet-Parker model.  (Any possible resistive dissipation in this pre-flare current sheet is neglected in our energy-release scenario.)  The two current sheet properties most significant for our model are the strength of the equilibrium magnetic field adjacent to it and the angle between the magnetic field it separates.

In addition to the magnetic properties, our model depends on the density of the 
flux tube plasma prior to its reconnection.  
This density determines both the Alfv\'en speed and the post-shock density.  We found that large observed emission measures demand fairly large densities, $n_{e,0}\approx8\times10^{10}\,{\rm cm}^{-3}$.  This demand is not unique to our model, since it derives from the limit on density enhancement at a single shock.  Indeed, our oblique shocks, similar to gas-dynamic shocks, produce a larger enhancement factor than switch-off shocks of two-dimensional models: $4$ {\em versus} $2.5$.  It thus seems impossible that the super-hot post-shock material could have arisen from pre-shock material at densities comparable to those observed in non-flaring active regions, $n_{\rm e}\approx10^9$.  Something in the flare process evidently ``pre-fills'' the flux --- at least the flux immediately adjacent to the current sheet.

While the high post-shock density was inferred from the observed emission measure, independent evidence for it comes from the occurrence of thermal bremsstrahlung from a plasma at $T\approx40$ MK confined to $\la10$ Mm along a magnetic field.  The mean-free paths of electrons at that temperature, $\ell_{\rm mfp}=100{\rm Mm}(10^{9}/n_{e,2})$, will greatly exceed the size of the source when density is low.  Low density plasma streams accelerated by field line retraction (\ie\ the RDs) would presumably pass through one another, rather than shocking.  Such counter-streaming has been observed in simulations of collisionless fast reconnection \cite{Drake2009}.  The fact that we observe bremsstrahlung from a Maxwellian electron population implies sufficient collisionality to thermalize the loop-top source: 
$\ell_{\rm mfp}\la 1$ Mm, and therefore very high post-shock density; 
$n_{e,2}\ga10^{11}\,{\rm cm}^{-3}$.

The pre-reconnection density $n_{e,0}$ is not, strictly speaking, a pre-flare density.  The thermal phase of the flare, to which our modeling has been devoted, began three minutes after the flare itself.  We have no estimate of the coronal density, even post-reconnection,
from the first three minutes (1:53:30 --- 1:56:40) where thick-target emission was the best fit.  It is entirely possible that the pre-flare density was far lower than the density we infer for flux tubes reconnected during the thermal phase of the flare.  Indeed, all observational evidence suggests pre-flare densities typical of ARs: $n_{\rm e}\approx 10^9$.  One possibility for large 
$n_{e,0}$ during that phase is that energy release during the initial non-thermal phase somehow enhanced the density of flux tubes in the vicinity of the current sheet 
{\em prior to} their reconnection.

We have not attempted to address the non-thermal phase with our purely fluid model, and therefore cannot include its energetic contribution in our estimate.  It is over this period that the inferred magnetic energy input diverges most from the energy output in \fig\ \ref{fig:net_erg}.  Our output rate neglects the precipitation of non-thermal particles, and uses conductive losses assuming collisional, Spitzer transport.  We therefore expect the input and output to disagree to some extent during this period, as they do.

In contemplating possible sources of non-thermal particles it is worth recalling the failure of fluid model when pre-reconnection densities are too low.  Low collisionality in the post-shock density leads to a non-thermal electron population with mean free paths comparable to, or larger than, the total loop length.  Furthermore, higher pre-reconnection Alfv\'en speeds lead to even higher post-shock temperature predictions, $T_{2}\approx v_{A,0}^2\approx n_{e,0}^{-1}$.  While these extremely large temperatures could not be collisionally thermalized within the loop, they are based on energy conservation laws which must still obtain.  We therefore expect a population of high energy, non-thermal electrons, accelerated directly by the 
Alfv\'enic loop retraction, the details of which awaits future kinetic calculations.

The above reasoning suggest that the non-thermal phase occurs when the pre-reconnection densities are low.  The ambient, pre-reconnection densities rise, perhaps due to evaporation driven by observed precipitation of non-thermal particles, and subsequent reconnection occurs at densities high enough to produce collisionally thermalized plasma at the loop-top. 

This evaporation scenario is similar to standard flare models except that ours requires evaporation to occur on a flux tube {\em before} that flux tube reconnects to release its energy.  It is the rapid energy release that raises the temperature of the already-dense material to $T\approx40$ MK.  Evaporation flow is unlikely to cross field lines so evaporative pre-filling seems to require a single flux tube to undergo multiple reconnections.  In fact, models of reconnection in more complex magnetic topologies do predict a single flux tube to reconnect several times, passing through different separators into intermediate domains 
\cite{Longcope2007,Longcope2007c}.  Moreover, we observe post-flare loops 
(\fig\ \ref{fig:tr_sum}) in domains, such as $P09$ -- $N01$, before we observe them in $P04$ -- $N01$, for which we compute the net flux transfer.  Sufficient time between successive reconnections would permit a tube to be ``pre-filled'' by evaporation driven by energy released by this preceding reconnection.

The requirement that evaporation pre-fill loops before their final, energy-releasing reconnection, offers insight into the rarity of thermal flares of the kind we have modeled. Evaporative pre-filling is probably easier, and therefore a more common occurrence, in small compact flares, with short loops, than in large eruptive flares.  The loops in this flare, ranging from $L=13$ -- 22 Mm (doubling the visible EUV lengths in \eq\ [\ref{eq:Lloop}]), are short compared to those in eruptive events.  Furthermore, the reconnection was estimated to occur very low in the corona: $\approx10$ Mm.  The emergence of new flux into this small active region appears to have loaded it with $\Delta W\approx 10^{32}$ ergs, which was later liberated by reconnection.  This produced short post-reconnection loops which were quickly filled even as they continued to be shortened.

Uncommon are flares whose energy release is sufficiently dominated by thermal emission to be modeled only as a fluid, however, we believe our energy release scenario may have broader applicability.  Large eruptive flares often exhibit supra-arcade downflows, which offer evidence for the same kind of post-reconnection retraction used in our model 
\cite{McKenzie1999,Sheeley2004,McKenzie2009}.  Ultimately any kind of magnetic reconnection releases energy by permitting field lines to become shorter (except in the special case of anti-parallel fields, where some of the field energy is annihilated).  If it is fast reconnection then the shortening will occur at the Alfv\'en speed, and plasma on the field lines will be compressed super-sonically.  We therefore expect the energy release scenario in our model to be fairly common, although perhaps not always treatable by fluid models when densities are too low.  Even in large, two-ribbon flares, however, the gradual phase shows little evidence of non-thermal particles and yet can account for substantial share of the energy release \cite{Emslie2005}.  Since it persists far longer than the cooling time of a single loop, the gradual phase must be maintained as a succession of new reconnection sites  create and energize new loops \cite{Warren2002,Reeves2002}.  The energy release in these cases may well occur through the same shock heating scenario introduced here, evidently without any non-thermal particles.

\bigskip

This work was supported under the NSF-REU program and by a joint NSF/{-}DOE grant.  We thank Jim Drake, Terry Forbes, Jack Gosling, Silvina Guidoni, Mark Linton and Harry Warren for helpful discussions during the preparation of the manuscript.  We also thank the anonymous referee for suggestions for improving the manuscript.


\end{document}